\documentclass[preprint]{aastex}
\usepackage{graphicx,amsmath}

\slugcomment{The Astrophysical Journal, accepted}

\ifx\ensuremath\relax\else\def\ensuremath#1{\ifmmode{#1}\else{$#1$}\fi}\fi

\newcommand{\scrbox}[1]{\ensuremath{{\mbox{\scriptsize #1}}}}

\newcommand{\teff}{{\ensuremath{T_{\scrbox{eff}}}}}
\newcommand{\Msol}{\ensuremath{\,\mbox{\it M}_{\odot}}}
\newcommand{\Mstar}{\ensuremath{\it \,M_{*}}}

\newcommand{\Kelvin}{\,\mbox{K}}
\newcommand{\MS}{main--sequence}
\newcommand{\gr}{\ensuremath{g_{\scrbox{rad}}}}
\newcommand{\tbcz}{{\ensuremath{T_{\scrbox{bcz}}}}}

\renewcommand{\H}{\mbox{H}}
\newcommand{\He}{\mbox{He}}
\newcommand{\B}{\mbox{B}}
\newcommand{\Fe}{\mbox{Fe}}

\newcommand{\Si}{\mbox{Si}}
\newcommand{\Cr}{\mbox{Cr}}

\newcommand{\F}{\mbox{F}}
\newcommand{\Ox}{\mbox{O}}

\newcommand{\Li}{\mbox{Li}}

\newcommand{\T}{_{\rm{T}}}

\begin{document}

\title{MODELS OF METAL POOR STARS WITH GRAVITATIONAL SETTLING AND RADIATIVE ACCELERATIONS: I. EVOLUTION AND ABUNDANCE ANOMALIES}

\author{\textsc{O. Richard\altaffilmark{1} and G. Michaud\altaffilmark{1} and  J. Richer\altaffilmark{1}} }
\affil{D\'epartement de physique, Universit\'e de Montr\'eal, Montr\'eal,
  PQ, Canada, H3C~3J7}
\email{richard@CERCA.UMontreal.CA, michaudg@CERCA.UMontreal.CA, jacques.richer@UMontreal.CA}
\and
\author{\textsc{S. Turcotte}}
\affil{Lawrence Livermore National Laboratory, 7000 East Ave., L-413, Livermore,
CA 94550}
\email{sturcotte@igpp.ucllnl.org}
\and
\author{\textsc{S. Turck-Chi\`eze}}
\affil{SAp/CEA 
Service d'Astrophysique du Commissariat \`a l'Energie Atomique, Saclay,
L'Orme des Merisiers, B\^atiment 709,
91191 Gif-sur-Yvette Cedex}
\email{cturck@cea.fr}
\and
\author{\textsc{Don A. VandenBerg}}
\affil{Department of Physics \& Astronomy, University of Victoria, 
       P.O.~Box 3055, Victoria, B.C., V8W~3P6, Canada}
\email{davb@uvvm.uvic.ca}

\altaffiltext{1}{CEntre de Recherche en Calcul Appliqu\'e (CERCA), 
                 5160~boul.~D\'ecarie, bureau~400,      
                 Montr\'eal,~PQ, Canada, H3X~2H9}
\newpage

\begin{abstract} 
Evolutionary models have been calculated for Pop II stars of 0.5 to 1.0$\Msol$ 
from the pre-\MS{} to the lower part of the giant branch. Rosseland opacities and radiative accelerations
were calculated taking into account the concentration variations of 28 chemical species, including 
all species contributing to Rosseland opacities in the OPAL tables.  The effects of 
radiative accelerations, thermal diffusion and gravitational settling  are included.  
While models were calculated both for $Z=0.00017 $ and $0.0017$, we concentrate on models with $Z=0.00017 $ 
in this paper.  
These are the first Pop II models calculated taking 
radiative acceleration into account. It is  shown that, at least in a 0.8 \Msol{} star, it is a better approximation 
not to let \Fe{} diffuse than to calculate its gravitational settling without including the effects of \gr(Fe).  

In the absence of any turbulence outside of convection zones, the effects of atomic
 diffusion are large mainly for stars more massive than 0.7$\Msol$. 
Overabundances are expected in \emph{some} stars with $\teff{} \ge 6000 $\Kelvin{}. 
Most chemical species heavier than CNO are affected.  At 12 Gyr, overabundance factors may reach 10 in some cases (e.g. for Al or Ni)
 while others are limited to 3 (e.g. for Fe).

The calculated surface abundances are
 compared to recent observations of abundances in globular clusters as well as
 to observations of Li in halo stars.  It is shown that, as in the case
 of Pop I stars, additional turbulence appears to be present.  

Series of models with different assumptions about the strength of turbulence were 
then calculated.  One series minimizes the spread on the Li plateau while another was chosen 
with turbulence similar to that present in AmFm stars of Pop I.  Even when turbulence
is adjusted to minimize the reduction of Li abundance, there remains a reduction by a factor of at least 1.6
from the original Li abundance. Independent of the degree of turbulence in the 
outer regions, gravitational settling of He in the central region reduces the 
lifetime of Pop II stars by 4 to 7 \% depending on the criterion used.  The effect on the age of the oldest clusters is discussed in 
an accompanying paper.

Just as in Pop I stars where only a fraction of stars, such as AmFm stars, have abundance anomalies, one should look for 
the possibility of abundance anomalies of metals in some Pop II turnoff stars, and not necessarily in all.
Expected abundance anomalies are calculated for 28 species and compared to observations of M92 
as well as to Li observations in halo field stars.

\end{abstract}


\section{Astrophysical context}
\label{sec:context}

Helioseismology has confirmed the importance of gravitational settling in the Sun's external regions (\citealt{GuzikCo92}; \citealt{GuzikCo93}; \citealt{ChristensenDalsgaardPrTh93}; 
 \citealt{Proffitt94}; \citealt{BahcallPiWa95}; \citealt{GuentherKiDe96}; \citealt{RichardVaChetal96}; \citealt{BrunTuZa99}).  Turnoff 
stars in globular clusters 
are only slightly less massive than the Sun and have convection zones that tend to be somewhat shallower.
In solar type stars, radiative
 accelerations have been shown to become equal to gravity for some metals around the end of the main sequence 
lives \citep{TurcotteRiMietal98}.  The 
question of the atomic diffusion of metals in Pop II stars then naturally arises.  It is currently of special 
interest because large telescopes 
are now making  possible the determination of the abundance of metals in the turnoff stars of globular clusters.  
In this paper evolutionary models that take into account the diffusion 
of He, LiBeB and metals in Pop II stars are presented for the first time.  Surface abundances may then be used as additional 
constraints 
in the determination of the age of globular clusters and of the Universe (\citealt{VandenBergRiMietal2001}, hereafter Paper II).  
Given their old age, Pop II stars are 
those where the slow effect of atomic diffusion  has 
the largest chance to play a role on the evolutionary properties. Previously published evolutionary models of Pop II stars have included some of the effects of diffusion 
(\citealt{DeliyannisDeKa90}; \citealt{ProffittMi91}; \citealt{ProffittVa91}; \citealt{SalarisGrWe2000}; \citealt{StringfellowBoNoetal83})
but never included the effects of the diffusion of metals with their radiative accelerations self consistently.

Determining constraints on stellar hydrodynamics from abundance observations requires knowing the original chemical composition of the star.  In 
Pop II stars there are larger variations in original abundances than in Pop I.  For that reason it is essential to use globular clusters 
for abundance determinations  of metals since only then does one have a handle on the original abundances  from observations of cluster giants.
However  LiBeB  have an  origin partly different from that of most metals.  Coupled with their sensitivity to low temperature nuclear burning, 
their abundance determination in halo stars provides useful constraints on hydrodynamics even if determinations in cluster stars
would be preferable. \footnote{Only in the case of NGC 6397 \citep{MolaroPa94} and M92 \citep{BoesgaardDeKietal99} have Li
abundances been determined for turnoff stars, and these data  are much less precise than those obtained
for halo field stars.  The M92 data  are discussed later in the paper. }

The observation by \citet{SpiteSp82} of a plateau in the Li concentration  over a relatively large \teff{} interval of Pop II
 stars has now been confirmed by many observations.  Furthermore, the Li concentration is constant while that of Fe
varies by more than a factor of 100 \citep{Cayrel98} from $[\Fe/\H] = - 3.7$ to $-1.5$.  This shows 
the primordial origin of Li.  Its preservation for such 
a time interval over such a wide \teff{} interval   seriously challenges  
our understanding of convection and other potential mixing processes in those stars \citep{MichaudFoBe84}.  If there were no mixing process outside
of convection zones, the surface Li abundance would vary with \teff{}: at small \teff{} because of nuclear reactions 
($^7\Li (p,\alpha)^4\He$) and at large \teff{} because of gravitational settling.  Extending convection zones by a simple turbulence model does not solve 
the problem since, if the extension is sufficient to reduce Li settling enough in the hotter stars,
 it causes excessive Li destruction in the cooler stars of the plateau \citep{ProffittMi91}.
Efforts were also made to link such an extension to differential rotation (\citealt{Vauclair88} or, as parametrized in the Yale
 models, \citealt{PinsonneaultWaStetal99})
 but the small apparent dispersion in the plateau makes this model unlikely. \citet{RyanNoBe99} even claim that  the 
destruction may not be by more than 0.1 dex.

Several groups (\citealt{CayrelSpSpetal99}, \citealt{HobbsTh94, HobbsTh97}, \citealt{SmithLaNi93}) have also observed $^6\Li$ in halo
 stars. Since $^6\Li$ is destroyed at a smaller \teff{} than $^7\Li$, its survival in old stars implies 
a strict and small upper limit on the amount of mixing in those stars.  The stars where a detection has been 
made are all concentrated close to the turnoff \citep{CayrelSpSpetal99}.  While a mechanism has been suggested to produce $^6\Li{}$ in 
the Sun \citep{RamatyTaThetal2000}, it is not expected to significantly affect atmospheric values so that a  star in which $^6\Li{}$ is seen in the
 atmosphere may not have destroyed a significant fraction of its original $^6\Li$.

 \citet{BoesgaardDeKietal99} determined Be abundances in halo stars as well as a linear correlation with the Fe abundance suggesting that 
Be has not been destroyed in those stars.

The chemical composition of globular clusters is attracting more attention as large 
telescopes make the determination of the surface chemical composition of turnoff stars
 possible.  The determination of the abundance of metals makes it 
possible to put additional constraints on the hydrodynamics of Pop II stars. Some observers have reported factors of 2 differences between Fe abundances
in red giants and subgiants \citep{KingStBoetal98}.
However \citet{RamirezCoBuetal2001} and \citet{GrattonBoBretal2001} find no difference between the \Fe{} abundances in red giants and turnoff stars.  \citet{TheveninChdeetal2001} find the same relative abundances as in the Sun in the turnoff stars of NGC$\,$6397.
Unexplained anticorrelations between the abundances of O and Na have been observed by \citet{GrattonBoBretal2001}.

In this paper, the surface abundances to be expected in Pop II stars are calculated under different
assumptions for the internal stellar hydrodynamics.  The first and simplest  assumption is that there is
no macroscopic motion outside of convection zones.  There remains only atomic diffusion, including the effects 
of gravitational settling, thermal diffusion and radiative accelerations, as a transport 
process outside of convection zones.
On the main sequence, such models have been shown to 
lead to larger abundance anomalies than are observed \citep{TurcotteRiMi98} but also to the appearance 
of an additional convection zone 
caused by the accumulation of iron peak elements \citep{RichardMiRi2001}.  
A relatively simple parametrization of turbulence, 
corresponding to an extension of those Fe convection zones by a factor of about 5 in mass, 
was shown to lead to a simple explanation of the AmFm phenomenon \citep{RicherMiTu2000}.  A similar parametrization is 
used here in Pop II stars and 
leads to our second series of models.
Finally, we use an additional parametrization that is chosen in order to minimize the reduction of the surface 
Li concentration, so as to provide the best representation of the Spite plateau.   \citet{ProffittMi91} introduced a similar parametrization of turbulence
to compete with He and Li settling (other parametrizations of turbulence were introduced for the Sun by  \citealt{RichardVaChetal96} and \citealt{BrunTuZa99}).   \citet{Basu97}
has shown that weak turbulence below the solar convection zone also improves agreement with the solar pulsation spectrum.
For comparison purposes, 
series of models are also calculated  
without diffusion and one model is calculated with gravitational settling but without \gr{} (see subsubsection~\ref{sec:structure}).

\section{CALCULATIONS}
\label{sec:calcul}

\subsection{Models}
\label{subsec:model}

The models were calculated as described in \citet{TurcotteRiMietal98}.
The radiative accelerations are from \citet{RicherMiRoetal98} with
the correction for redistribution from \citet{GonzalezLeAretal95} and
\citet{LeblancMiRi2000}. The atomic diffusion coefficients were taken from \citet{PaquettePeFoetal86}.
 The uncertainties in the atomic and thermal diffusion coefficients
have been discussed by \citet{MichaudPr93}, and by \citet{Michaud91} in particular for the temperature--density 
domain of the interior of Pop II stars and the Sun.

Turbulent transport is included as in \citet{RicherMiTu2000} and \citet{RichardMiRi2001}
where the parameters specifying turbulent transport coefficients are
indicated in the name assigned to the model.  For instance, in the T5.5D400-3 model,
 the turbulent diffusion coefficient, $D\T$,  is 400 times larger
than the He atomic diffusion coefficient at $\log T = 5.5$ and varying as $\rho^{-3}$. To simplify writing, T5.5 will also be used
instead of T5.5D400-3 since all models discussed in this paper have the D400-3 parametrization.  The $\rho^{-3}$ dependence is suggested (see 
\citealt{ProffittMi91a})
by observations that the Be solar abundance today is hardly smaller than the original Be abundance (see for instance \citealt{BellBaBa2001})

All models considered here were assumed to be chemically homogeneous on the
pre-main sequence with the abundance mix appropriate for Pop II stars.  The relative 
concentrations used are defined in Table~\ref{tab:Xinit}. The relative concentrations of the \emph{alpha}  
elements are increased compared to the solar mix as is believed to be appropriate in Pop II stars \citep{VandenBergSwRoetal2000}.

\subsection{Mixing length}
\label{subsec:alpha}
In \citet{TurcotteRiMietal98}, the solar luminosity and radius at the solar age were used 
to determine the value of $\alpha{}$, the ratio of the mixing length to pressure scale height, 
and of the He concentration in the zero age Sun.   They calibrated $\alpha{}$ both in models using 
 Krishna-Swamy's $T(\tau){}$ relation \citep{Krishna_Swamy66} (both with and without atomic diffusion) and in models 
using Eddington's  $T(\tau){}$ relation (only with diffusion).
 The 
He concentration mainly affects the luminosity while $\alpha{}$ mainly determines the radius, 
through the depth of the surface convection zone. The required 
value of $\alpha{}$ was found to be slightly larger in the diffusive than in the non diffusive
models because an increased value of  $\alpha{}$ is needed to compensate for He
and metals settling from the surface convection zone.   The increased $\alpha{}$ in the 
diffusion models of the Sun is then
 determined by the settling occurring immediately below the solar surface convection zone.
 See \citet{FreytagSa99} for a discussion of uncertainties
related with the use of the mixing length in Pop II stars.  

  Some of the  models presented in this paper  were calculated with  
turbulence below the convection zone.  The appropriate value of $\alpha{}$ to use then depends 
on the depth of  turbulence.  If turbulence is large enough to eliminate settling at the depth of
solar models, the most appropriate  $\alpha{}$ to use is that determined for non diffusing
 solar models.  If the adopted turbulence is more superficial than the solar convection zone,
the appropriate  $\alpha{}$ to use is that for the solar models with diffusion. Given the uncertainty, it was chosen to use the same value 
of $\alpha$ for all models, both with and without diffusion, calculated with  Eddington's  $T(\tau){}$ relation.   In order to
give an estimate of the uncertainty related to  $\alpha{}$, two  values of  $\alpha{}$ (one determined by the solar model
without diffusion and one by the solar model with diffusion) will be
used for one series of models with diffusion (see subsection~\ref{sec:alternate_models}).

The calculated series of models are identified in Table~\ref{tab:parameters}. The series labeled KS$\alpha{}$ used a value of $\alpha{}$ determined 
from a solar model with diffusion while the series labeled KS used a value of $\alpha{}$  determined 
from a solar model without diffusion.

\section{EVOLUTIONARY MODELS}
\label{sec:EVOLUTION}
Four series of evolutionary models were calculated.   In the first subsection, the models with 
atomic diffusion are presented.  In the second one, the models with turbulence and those without diffusion 
are introduced and compared to the models with atomic diffusion.  Two series of models with turbulence are discussed
in some detail: a series that minimizes Li underabundance and one that contains a level of turbulence similar to that needed to reproduce
the observations of AmFm stars.  In the 
last subsection, the effect of changing boundary conditions is analyzed.

\subsection{Evolution with atomic diffusion}
\label{sec:with diffusion}

The \teff{} and $\log{}g$ as a function of age as well as the luminosity as a function of \teff{} are shown in Figure~\ref{fig:historic_noturb}
for evolutionary models calculated including atomic diffusion and  only
 those physical processes that can currently be evaluated properly from first principles. 
That figure also contains  the time variation of  the depth of the 
surface convection zone, of the temperature at its bottom and of central H concentration.  
Models were calculated for 0.5 to 1.0 \Msol{} stars from the pre-\MS{} to the bottom of the giant branch except 
 for the 0.85, 0.9 and 1.0 \Msol{} stars which were stopped earlier  because of numerical instabilities.
All models shown have $Z = 0.00017$.  The mass interval was chosen to minimize the possibility of interpolation 
errors in the construction of isochrones (see Paper II).

The variation of the mass in the surface convection zone is large enough to play a major role on surface
 abundance evolution.
In the next two subsections,  the effect of radiative accelerations on surface abundances are related to the regression
of the surface convection zone during evolution.

\subsubsection{Radiative accelerations}

The radiative accelerations below the surface convection zone play the most important role in 
determining the surface abundance variations
(see Fig.~\ref{fig:grad}).  As the evolution of a 0.8 \Msol{} star proceeds during the main sequence phase, 
the surface convection zone 
becomes progressively thinner: between 6 and 11 Gyr, the mass in the convection zone is reduced by a factor of 
about 20, from $\log \Delta M/\Mstar = -3.0$ to $-4.3$ 
(see  Fig.~\ref{fig:historic_noturb}).  The effect of this regression of the surface convection zone
 may be seen in the surface abundances through the dependence of \gr{} on nuclear charge.  The \gr{} on Li
has a maximum at $\log \Delta M/\Mstar \simeq  -5$ where it is in the hydrogenic configuration.  It is 
completely ionized deeper in the star so that
  \gr(Li){} is progressively reduced deeper in.  The chemical species with a larger nuclear charge become 
ionized deeper in so that they are 
in an hydrogenic configuration deeper in and have
the related maximum of their \gr{} at greater depths.  For \B{}, it is at $\log \Delta M/\Mstar \simeq  -4$, for \Ox{}
 at $\log \Delta M/\Mstar \simeq  -2$ and for 
\Si{} at $\log \Delta M/\Mstar \simeq  -1$.  Another maximum in the \gr{} appears when a species is in between the atomic configurations 
of \Li{} and \F.  This maximum occurs at $\log \Delta M/\Mstar \simeq  -4$ for P, and $-2$ for \Cr{}.  These maxima may be followed on 
the figure for other species.  As may be seen from  Figure~\ref{fig:grad}, when the atomic species is in the hydrogenic configuration, the \gr{} 
is rarely larger than gravity.  It is larger only for Be, B and C.  At the maximum of \gr{} between the \Li{}-like and \F-like configurations, 
however, \gr{} is usually larger than gravity.  Contrary to what happens in Pop I stars \citep{RicherMiRoetal98}, the \gr{} do not depend
 on the abundance of the species
since it is too small to cause flux saturation.

\subsubsection{Abundances}
The time variation of surface abundances in Pop II stars with atomic diffusion is shown in Figure~\ref{fig:surf_abundance_diff}.  The surface abundances reflect 
the variation of the \gr{} below the surface convection zone as it moves toward the surface.  From Figure~\ref{fig:historic_noturb}, one sees 
that the depth of the surface convection zone gets progressively smaller during evolution until hydrogen is exhausted from the center.  In the 0.8 \Msol{} model, it starts 
at a depth of  $\log \Delta M/\Mstar \simeq  -2$ and is at $\log \Delta M/\Mstar \simeq  -4.5$ after 11 Gyr.  One sees in Figure~\ref{fig:grad} 
that \gr(Fe) is smaller than gravity when the convection zone is deep but larger than gravity when it is more superficial.  In the early evolution, Fe
 settles gravitationally though \emph{less rapidly} than He or C, largely because \gr(Fe) is not much smaller than gravity.  Around 9 Gyr, \gr(Fe) 
becomes larger than gravity below the surface convection zone. 
 One sees the effect in the surface abundances (Figure~\ref{fig:surf_abundance_diff}) 
where the surface Fe concentration 
starts increasing around 9 Gyr, becoming larger than the original concentration around 10 Gyr and reaching an 
overabundance by a factor of 
about 5 just before the star becomes a subgiant, at which point the Fe abundance goes back to 
very nearly its original value as the convection zone becomes more massive.  Similar
 remarks apply to other chemical species.  

The depth dependence of the abundance variations is shown for a  0.8 \Msol{} model in Figure~\ref{fig:intern_abundances}.  One first notices that the
abundance of $^6\Li$ is modified by nuclear reactions over the inner 98\,\% of the mass.  Atomic diffusion modifies the abundances over the outer 1\,\% of 
the mass.  This leaves hardly any intermediate zone. At 6.1 Gyr, the surface convection zone extends down to $\log \Delta M/\Mstar \simeq  -3.0$,  
where the \gr{} are smaller than gravity for most species.  Only K, Ca and Ti  start to be supported: their abundances in the surface convection zone are larger than 
immediately below.  As evolution proceeds, the convection zone recedes. 
 After 9.1 Gyr, atomic species between Na and Ni are supported as the \gr{} is larger than gravity below 
the bottom of the convection zone. This continues as the evolution proceeds.  After 11.4 Gyr, the Fe abundance, for instance, is 
3 times larger than the original abundance in the surface convection zone.  That increased concentration of Fe is caused by the migration
of Fe above $\log \Delta M/\Mstar \simeq  -3.5$ where its \gr{} becomes larger than gravity.  Between $\log \Delta M/\Mstar \simeq  -4.5$ and $-2$, the 
concentration of Fe decreases since it is either pushed into the surface convection zone by radiative accelerations 
(for $\log \Delta M/\Mstar \geq  -3.5$) or settles gravitationally below.  On the scale of that figure, the concentration variations of Fe are not seen 
for $\log \Delta M/\Mstar \geq  -2$.  It sinks toward the center of the star where a small Fe overabundance  appears.  The diffusion time scales
are long enough that the effects amount to nearly a 10\,\% increase at the center after 11.4 Gyr (see Fig.~\ref{fig:center_zoom}).  This figure may be 
compared to Figures 15 and 16 of \citet{TurcotteRiMietal98} where the increase in Fe concentration at the center is by about 3\,\% at the age of the Sun.
Transformation of C and O into N also modifies the  value of $Z$ so that it increases by only  4\,\%  at the center and has a maximum outside of the 
region where O is transformed into N.

In stars of other masses, surface abundances are also mainly determined by the depth reached by the surface convection zone.  
In the 0.7 \Msol{} model, the surface convection zone never gets thinner than $\log \Delta M/\Mstar \simeq  -3$ so that no overabundances
 appear during evolution and underabundances are limited to  a factor of about 2.  In the 0.9\Msol{} model on the other hand, the surface Fe 
 abundance starts increasing around 2 Gyr, when the surface convection zone becomes thinner than $\log \Delta M/\Mstar \simeq  -4$, 
 where, as seen above, \gr(Fe) becomes larger than gravity.  In this star, 
Fe becomes overabundant around 3 Gyr. In  Pop II stars of larger mass, the overabundances are larger because the surface convection zones are thinner.

\subsection{Evolution with turbulent diffusion}
\label{sec:turbulent diffusion}

\subsubsection{Turbulent diffusion coefficient and Li abundance}

From Figure~\ref{fig:historic_noturb}, one sees that even in one single model, say the 0.8 \Msol{} one, the mass in the convection 
zone varies from about $\log \Delta M/\Mstar \simeq  -2.1$ to $-4.2$ during the \MS{} evolution.  Linking 
turbulence to the position of
 the convection zone could never produce a small variation of Li: if one increased the depth of the convection zone
 by a fixed factor, large enough to eliminate
 settling around 11 Gyr for instance (a factor of 100 or so is needed), this would lead to complete destruction of
 Li in the early evolution.  To minimize Li abundance reduction, it is essential, if one uses time independent parametrization,
 to link turbulence to a fixed $T$ and not to the bottom of 
convection zones.

The turbulent transport coefficients used in these calculations are shown in Figure~\ref{fig:coefficients}.  We chose to introduce 
the turbulence that  minimizes the reduction of the surface Li abundance.  We consequently defined the turbulent diffusion coefficient
  as a function of $T$ in order to  adjust it 
most closely to the profile that limits gravitational settling of Li while not burning Li.  The nuclear reaction $^7\Li(p,\alpha)^4\He$ is
 highly $T$ sensitive and the Li burning occurs  at $\log T \simeq 6.4$ (see \citealt{LumerFoAr90} for a detailed discussion).  Turbulent diffusion was adjusted to be smaller than atomic diffusion 
slightly below that $T$ so that turbulence would reduce settling as much as is 
possible in surface layers without forcing Li to 
diffuse by turbulence to $\log T \simeq 6.4$.

A  few series of evolutionary models were run in order to optimize turbulence parameters.  Examples of the resultant interior 
Li profiles are shown in Figure~\ref{fig:coefficients}.  The Li concentration always goes down rapidly as a function of increasing $T$ 
for $\log T \simeq 6.4$.  In the absence of turbulence, there is a peak in the Li concentration at $\log T \simeq 6.3$.  This peak 
disappears as the strength of turbulence is increased.  Turbulence in the T5.5D400-3 model is too weak to influence Li concentration 
in the $T$ interval shown.  Turbulence in the T6.0D400-3 model eliminates the Li abundance gradient caused by gravitational settling down to
$\log T \simeq 6.1$ while causing little reduction on the largest Li abundance at $\log T \simeq 6.3$.  Increasing turbulence further reduces 
the Li peak, and in the 
T6.2 model the surface Li abundance is smaller than in the T6.13 model, showing that one has passed the optimal value of turbulence.

It is interesting to note that the turbulent diffusion coefficient of the T6.09D400-3 model of 0.8 \Msol{} 
is a factor of about 
10 smaller than the turbulent diffusion coefficient found necessary to reproduce the solar Li surface abundance 
by \citet{ProffittMi91a}.  The latter corresponds approximately to the T6.2D400-3 turbulent
 diffusion coefficient profile.  
On the other hand the  T5.5D400-3 turbulent diffusion parametrization  corresponds approximately, as a function of $T$, 
to that found by \citet{RicherMiTu2000} to lead 
to the abundance anomalies observed in AmFm stars.

\subsubsection{Diffusion and stellar structure}
\label{sec:structure}
A comparison of the 0.8 \Msol{} models with and without diffusion is shown in Figure~\ref{fig:with_without}.  
First compare the Hertzsprung-Russell
 diagrams.  All models with diffusion (both with and without turbulence) nearly have the same evolutionary tracks 
(those with turbulence are somewhat hotter).  
However the one without atomic diffusion goes  to 
significantly higher \teff{} than all models with diffusion.  We have also verified that the turnoff temperature 
is reached 6 to 7\,\% earlier in the models with diffusion (with or without turbulence)
 than in the model without diffusion. 
These differences are reflected in the isochrones based on these tracks and therefore on  
globular cluster  age determinations (see Paper II).   They partially 
come from the structural effects of the global redistribution of He but also from the 4 to 5\,\% difference
in the age at which the central H abundance is exhausted.  This in turn is reflected in the shorter age 
(also 4 to 5\,\%) at which all 
diffusion models become 
subgiants, as seen in the rapid decrease of \teff{} but the rapid increases of the mass and $T$ at the 
bottom of the surface convection zone. 

While there are significant differences between the model without diffusion and all those with diffusion, the 
 effect of  the different turbulent transport models on the depth and mass at the bottom
 of the surface convection zone, and the time variation 
of the central H concentration are seen from Figure~\ref{fig:with_without} to be practically 
negligible on the scale of that figure.  Of the macroscopic properties defining 
the evolution of a 0.8\Msol{} star, only the \teff{} and $\log g$ appear to be modified by the turbulence models that were introduced.  
While it is not shown for
 other masses, it is also true for them.  However it is  shown in Paper II that the differences in \teff{} 
that seem small  on the scale of Figure~\ref{fig:with_without} are important in the comparison of 
calculated and observed isochrones. 
 
Given that there are, in the literature, calculations of Pop II evolutionary tracks with gravitational 
settling but no \gr{}, it is worth  noting that the track ($L-\teff{}$
relation; upper left hand part of Fig.~\ref{fig:with_without}) of such a model is, around turnoff, slightly cooler than the track of the model
 with diffusion and \gr{} but no turbulence.  
Furthermore the \Fe{} abundance is very different, by a factor of about 1000, at 12 Gyr. The \tbcz{} is 
smaller by 0.1 dex mainly because of the reduction of opacity caused by the reduction of the \Fe{} abundance.   \emph{It is  a better approximation 
not to let \Fe{} diffuse at all in a 0.8\Msol{} star than to calculate its gravitational settling without including the effects of \gr(Fe)}. 
The \Li{} abundance is  smaller by a factor of 2 in this case both because of the reduction of the \tbcz{} and because  \gr(Li) has some 
effect in the 0.8 \Msol{} model.

\subsubsection{Stellar structure and abundances}

The surface abundance anomalies caused by atomic diffusion are much more affected by turbulence than the
 macroscopic properties.
In the T5.5 model, turbulence reduces surface abundance anomalies but only in the time interval from 9 to 12 Gyr.  
Before 9 Gyr, the surface convection zone 
is deeper than  $\log T \simeq 5.7$ while in the  T5.5D400-3 model, the turbulent diffusion coefficient is larger than atomic diffusion only
for temperatures up to $\log T \simeq 5.8$.  Only when the surface convection zone has retracted sufficiently for turbulent diffusion to be 
larger than atomic diffusion below the convection zone, does turbulence have any effect. 
 In this model, the turbulent diffusion coefficient does not suppress completely the effects of atomic diffusion 
 (see Fig.~\ref{fig:with_without}).  It reduces them by an amount dependent on the chemical species.  
An Fe overabundance appears in the absence of turbulence but it is eliminated in the T5.5 model while the He 
underabundance is much less affected.  The  T6.0 model has the same turbulence profile as the  T5.5D400-3 model except that the profile is shifted towards higher temperatures by 
a factor of 3 in $T$ (see Fig.~\ref{fig:coefficients}); this leads to a much larger effect on abundance anomalies. It limits the effect of atomic diffusion on the surface abundances of He and Li to a factor of $\simeq{} 1.6$. 

The effect of turbulent transport on the interior concentrations in a 0.8 $\Msol${} star is shown, 
for two values of turbulence, in 
Figure~\ref{fig:abtot_dif_6.09}.  The no turbulence 
case is also shown for comparative purposes.  The internal 
concentration profiles are shown at 
nearly the same age in the three models.  Such a star  today would  be close to the turnoff.
It is also the 
evolutionary epoch when the effect of radiative accelerations and diffusion are largest and when the effect of 
adding a turbulent diffusion coefficient is greatest.
 In the \mbox{T6.09D400-3} case, the \gr{} never play a large role since turbulence mixes 
down to  $\log \Delta M/\Mstar = -2.3$ and the \gr{} are not much greater than gravity at that depth.  
In fact they are greater than gravity only for Ti, Cr, Mn and Fe and only around the end of evolution 
when \gr{} causes settling to 
slow down in the exterior region.  A  minimum in abundances appears 
 at  $\log \Delta M/\Mstar = -1.8$ because of that slow down in the settling from above 
while settling continues unabated below.
In the star with the T5.5D400-3 parametrization, there is mixing from the surface down to $\log \Delta M/\Mstar \simeq -4$.
Deeper in the star, radiative accelerations (see Fig.~\ref{fig:grad}) are greater than gravity over a sufficient
 mass interval to cause overabundances in the surface regions for chemical species between Al and Ca and for Ni.  
The \gr{} for Mg, Ti, Cr, Mn and Fe are just sufficient to bring them back to their original abundance while
 most of the chemical species between He and Na are less supported and remain underabundant in the atmosphere.  The 
exceptions are B and C which are very nearly normal.  This behavior is to be contrasted to that in the star with no
turbulence, where mixing by the surface convection zone extends from the surface down to $\log \Delta M/\Mstar \simeq -4.4$; the 
concentration gradients below the convection zone are very steep because of the absence of turbulence;
this leads to larger overabundances, in particular a factor of 10 for Si and S but of 3 for most Fe peak elements;  
while C becomes underabundant, B becomes overabundant.  Turbulence changes the position in the star where 
atomic diffusion dominates and so where the 
sign of $(\gr - g)$ matters.  However, turbulent transport also modifies the steepness of concentration gradients.
  Together with evolutionary time scales, these determine over-- vs under--abundances. This
shows the need for complete evolutionary models in order to determine surface abundances.

\subsection{Surface boundary conditions }
\label{sec:alternate_models}
Changing the surface boundary conditions has a significant effect on the depth of the surface convection zone
 (see Section~\ref{subsec:alpha} and Fig.~\ref{fig:M_Tbzc}).  
Two boundary conditions were used. Both are often
used for Pop\,II models.  In their solar 
models, \citet{TurcotteRiMietal98} used mainly Krishna Swamy's $T(\tau){}$ relation. 
 It is used 
here for some series of models.  Eddington's $T(\tau){}$ relation is also used: it was used for A and 
F Pop I stars by \citet{TurcotteRiMi98}  and \citet{RicherMiTu2000}.   In \citet{TurcotteRiMietal98}, the Sun is used to determine the 
appropriate value of $\alpha$ to use with each of these boundary conditions.  One may view the 
difference in convection zone mass between the models (Fig.~\ref{fig:M_Tbzc}) as a reasonable 
estimate of the uncertainty. During evolution, the uncertainty varies from a factor of 1.3 to 1.6 in mass.  This is 
sufficient to modify the fits to globular cluster isochrones (see Paper\,II).  It has however a less
pronounced effect on surface abundances than the uncertainty on turbulence.  It modifies significantly surface abundances 
only when no turbulence is included in the models.

\section{Surface abundances and observations}
\label{sec:surface}
The surface concentrations of the 28 species impose constraints on stellar models.  They will be 
shown for three series of models, the atomic diffusion models with no turbulence, those that mimic the turbulence
of AmFm stars and those that minimize the effects of transport on surface Li concentration.  Surface abundance 
variations as a function of time have been discussed above (see subsections~\ref{sec:with diffusion} 
and ~\ref{sec:turbulent diffusion}).  
They are discussed below as a function 
of atomic number and of $\teff$ in order to facilitate comparison to observations.  In the first subsection,
 we present the results for all calculated species for individual stars.  In the second subsection, we present results 
at a given age for stars of various masses as a function of $\teff$.  They are then compared to observations 
in the last subsection.

\subsection{As a function of atomic number}
\label{sec:atomic}
\begin{figure}
\centerline{
\includegraphics[width=\textwidth]{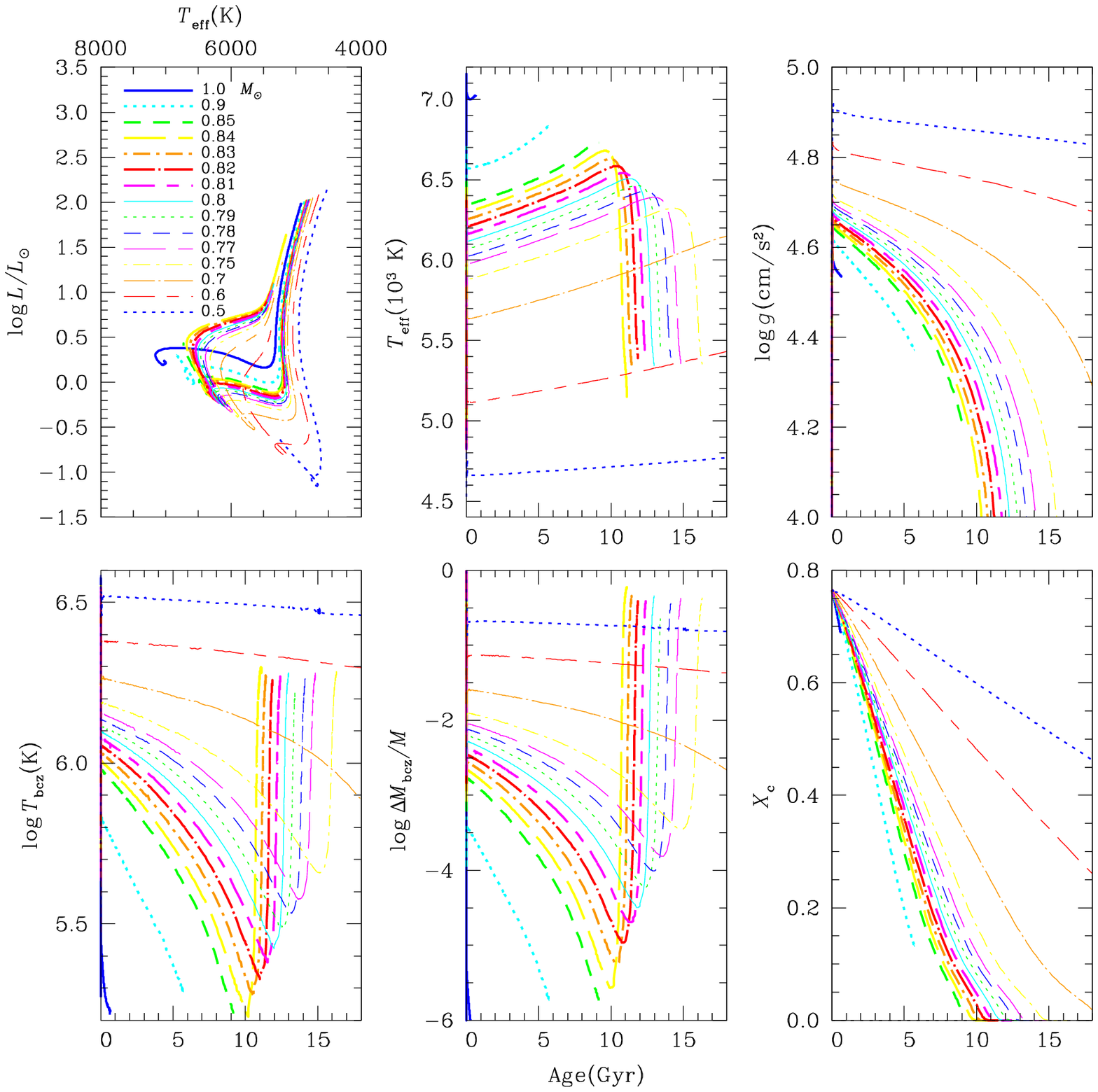}}
\caption{
Hertzsprung-Russell diagram, and time evolution of \teff{}, $\log g$, temperature at the base of the surface convection zone (\tbcz{}), 
mass above the base of the surface convection zone {\ensuremath{M_{\scrbox{bcz}}}} and mass fraction of hydrogen ($X_c$) at the  center of  stars
of 0.5 to 1.0 \Msol{} with $Z$ = 0.00017, or $ [\Fe/\H]=-2.31.$  All models were calculated with atomic
 diffusion and radiative accelerations but  no turbulent transport.  In the black and white figure, are
shown 6 of the 15  models  that may be seen in the color figure available in
 the electronic version of the paper.}
\label{fig:historic_noturb}
\end{figure}

\begin{figure}
\centerline{
\includegraphics[width=\textwidth]{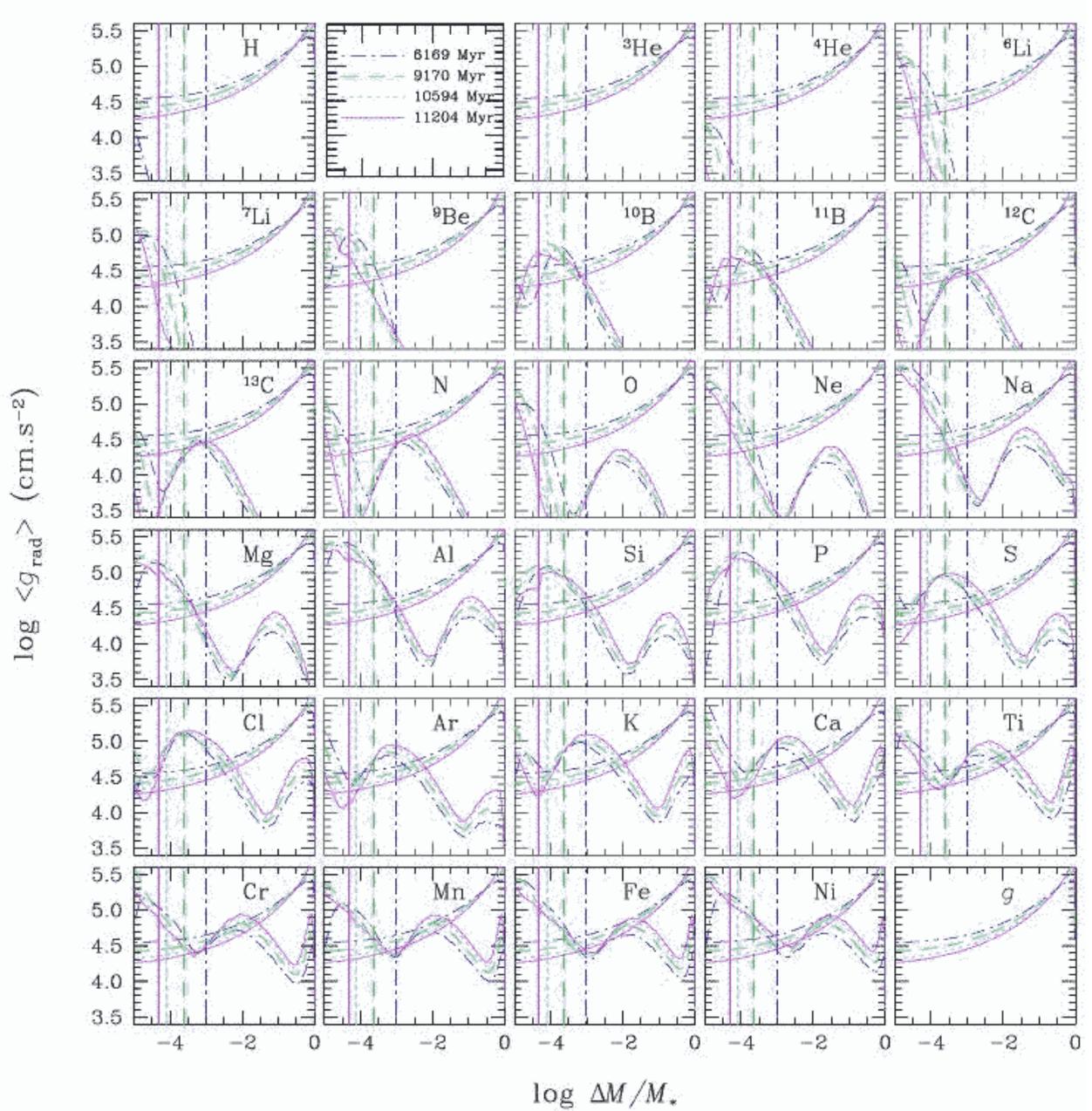}}
\caption{
Radiative accelerations in a Pop II star of 0.8 \Msol{} with $[\Fe/\H]=-2.31$ at four epochs  identified in the upper
part of the figure.  The vertical lines give the position of the bottom of the 
surface convection zone.  For other stellar masses, it is the position of the bottom of the convection zone that is most different.
The \gr{} of the various species varies from star to star but much less than the depth of convection zones.  
Gravity is shown in the lower right hand corner and repeated in each panel of the figure.  }
\label{fig:grad}
\end{figure}

\begin{figure}
\centerline{
\includegraphics[width=\textwidth]{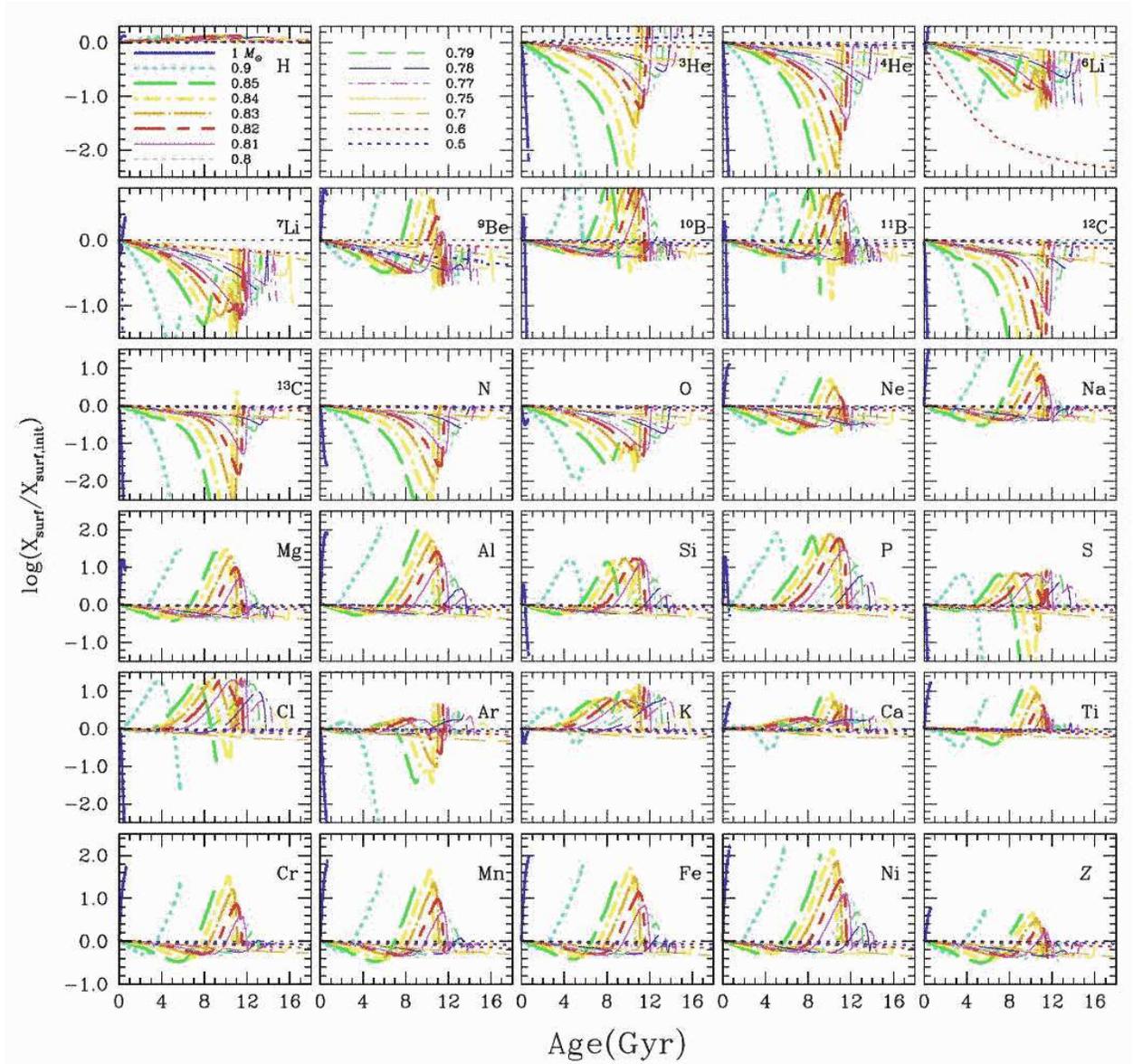}}
\caption{
Surface abundance variations in Pop II stars of 0.5 to 1.0 \Msol{} with $[\Fe/\H]=-2.31$. All models were 
calculated with atomic
 diffusion and radiative accelerations but  no turbulent transport.  In the black and white figure, are
shown 6 of the 15  models  that may be seen in the color figure available in
 the electronic version of the paper.  As is evident in the color version,  a vertical line drawn in
each panel of the figure, corresponding to a fixed age, permits one to evaluate the range of surface abundances
 of a species at that age (in, e. g., a globular cluster). }
\label{fig:surf_abundance_diff}
\end{figure}

\begin{figure}
\centerline{
\includegraphics[width=\textwidth]{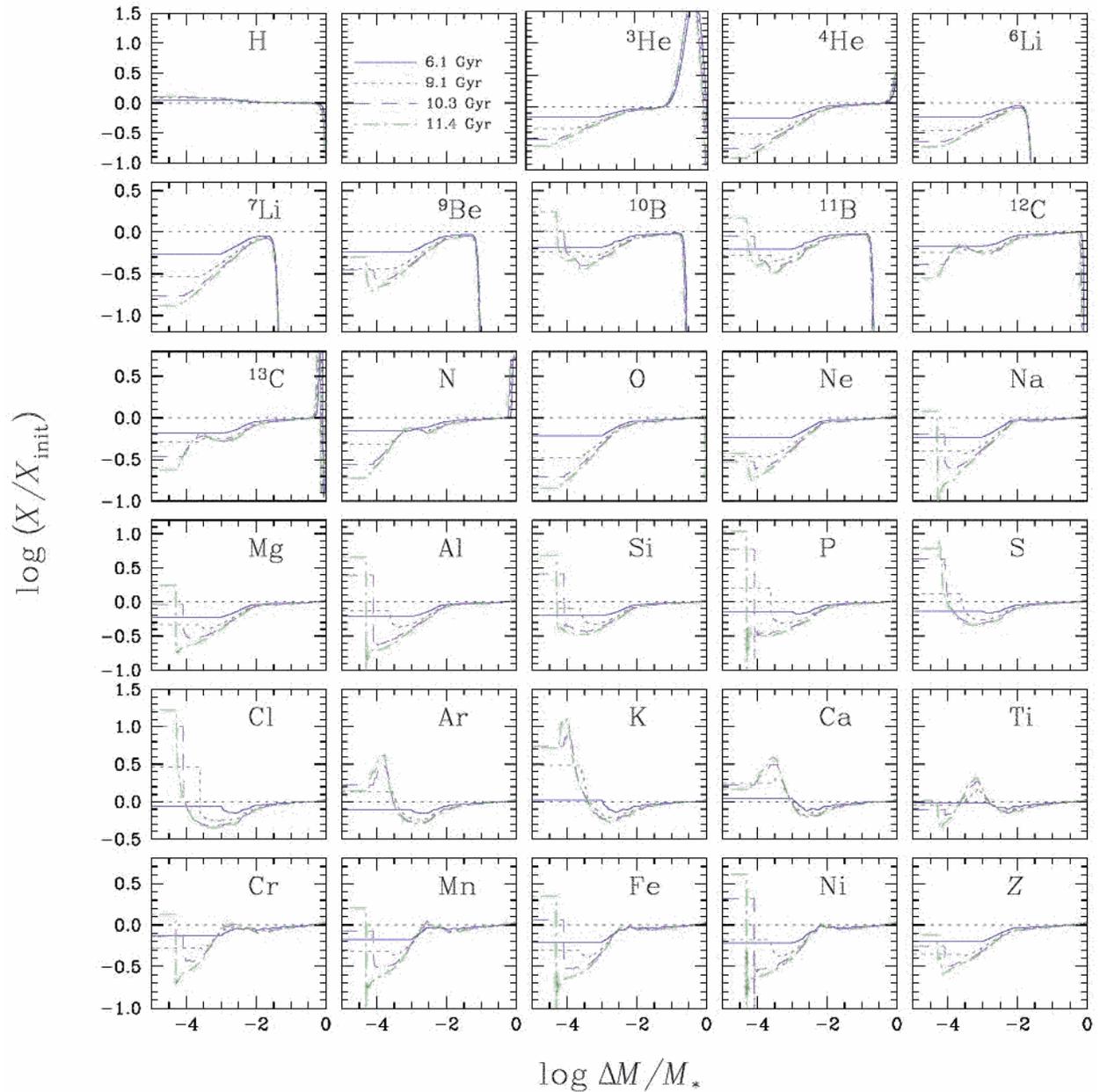}}
\caption{
Internal  abundance variations in a Pop II star of 0.8 \Msol{} with $[\Fe/\H]=-2.31$.  The profiles are shown at four different ages, 6.1,
9.1, 10.3 and 11.4 Gyr.  The last is shortly before the star moves to the giant branch. 
Calculations included  atomic
 diffusion and radiative accelerations but  no turbulent transport. }
\label{fig:intern_abundances}
\end{figure}

\begin{figure}
\centerline{
\includegraphics[width=\textwidth]{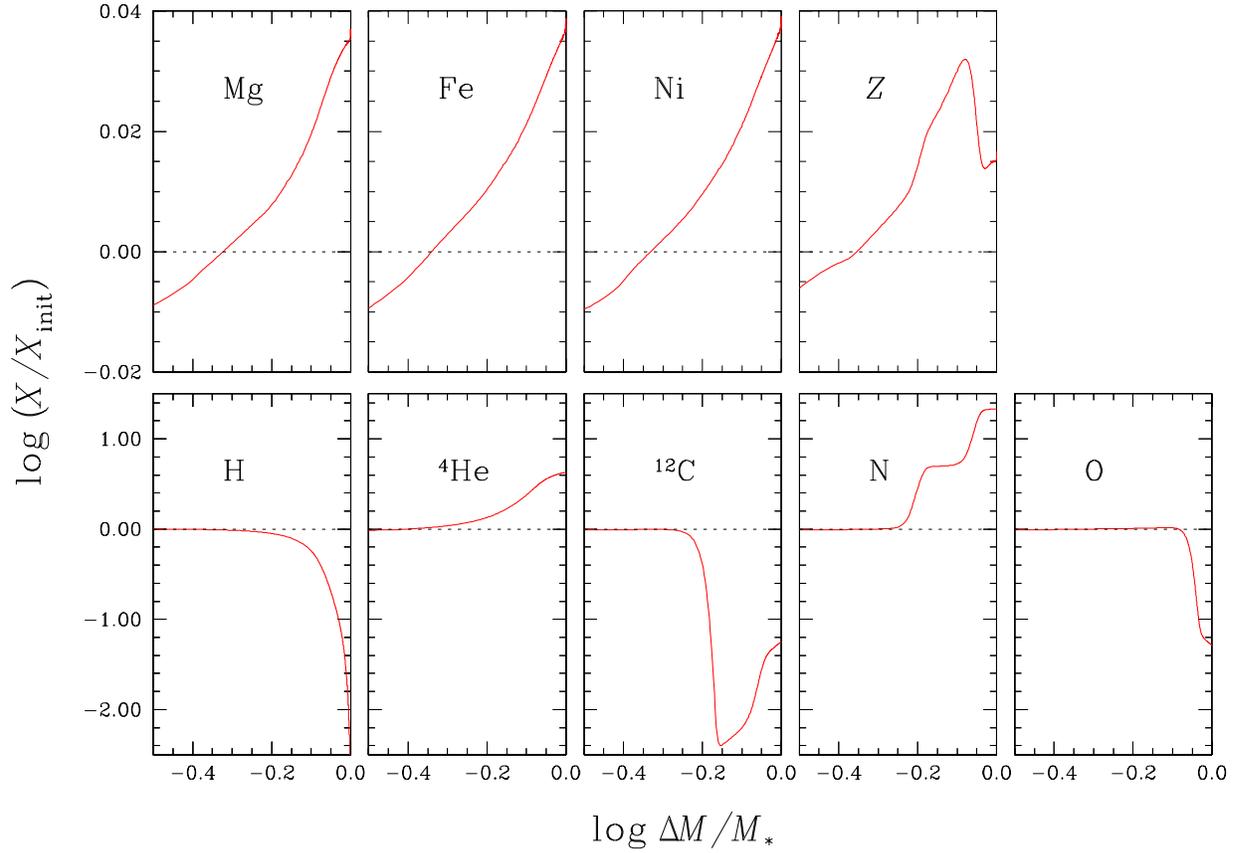}}
\caption{
Abundance variations in a Pop II star of 0.8 \Msol{} with $[\Fe/\H]=-2.31$ over the inner 2/3 of the mass of the star at an age of 11.8 Gyr.  Calculations included  atomic
 diffusion and radiative accelerations but  no turbulent transport. 
The zero abundance change occurs around 1/2 the mass of the star at  $\log \Delta M/\Mstar \simeq  -0.3$. 
 Interior to that point, the abundance 
of metals is generally larger than the original abundance.  The variations of CNO are due to their
 transformation by the CNO chain which would 
lead to a decrease of Z at the center if it were not for the settling of the other metals.  Only Mg, Fe and Ni are shown from the 
16 species between Ne and Ni included in the calculations but all others have very similar behavior.}
\label{fig:center_zoom}
\end{figure}

\begin{figure}
\centerline{
\includegraphics[width=\textwidth]{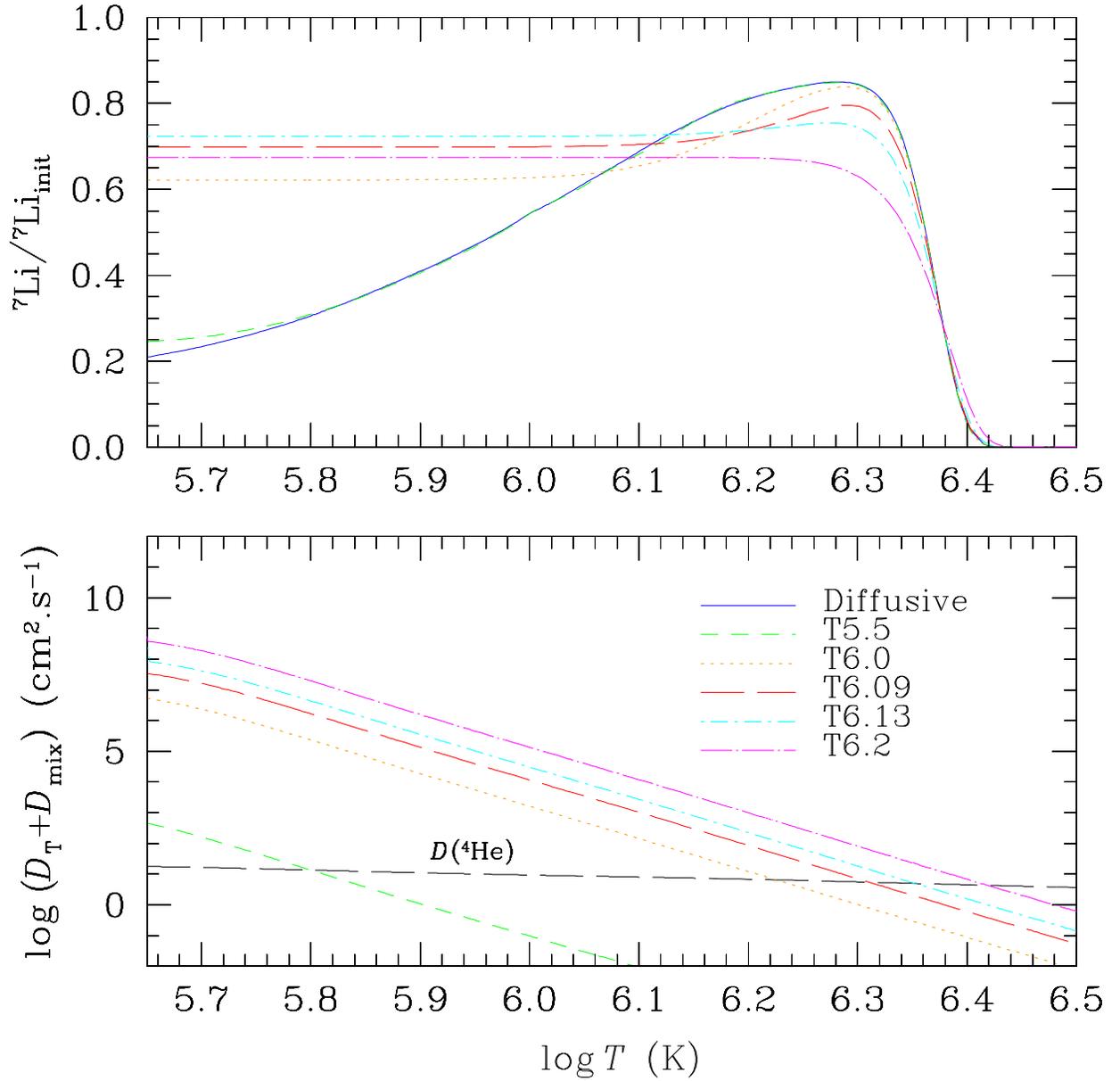}}
\caption{Temperature dependence of atomic and turbulent diffusion coefficients 
in 0.8 \Msol{} models.   The nearly horizontal line is the He atomic diffusion coefficient.  The other lines are various parametrizations of turbulence.  
In the upper part of the figure, the corresponding Li concentrations are shown, at an age of 10.2 Gyr, with the same line identification to give the link between 
turbulent transport and Li burning at $\log T \simeq 6.4$.  The parameters specifying turbulent transport coefficients are
indicated in the name assigned to the model.  For instance, in the T5.5D400-3 model,
 the turbulent diffusion coefficient, $D\T$,  is 400 times larger
than the He atomic diffusion coefficient at $\log T = 5.5$ and varying as $\rho^{-3}$.
}

\label{fig:coefficients}
\end{figure}

\begin{figure}
\centerline{
\includegraphics[width=\textwidth]{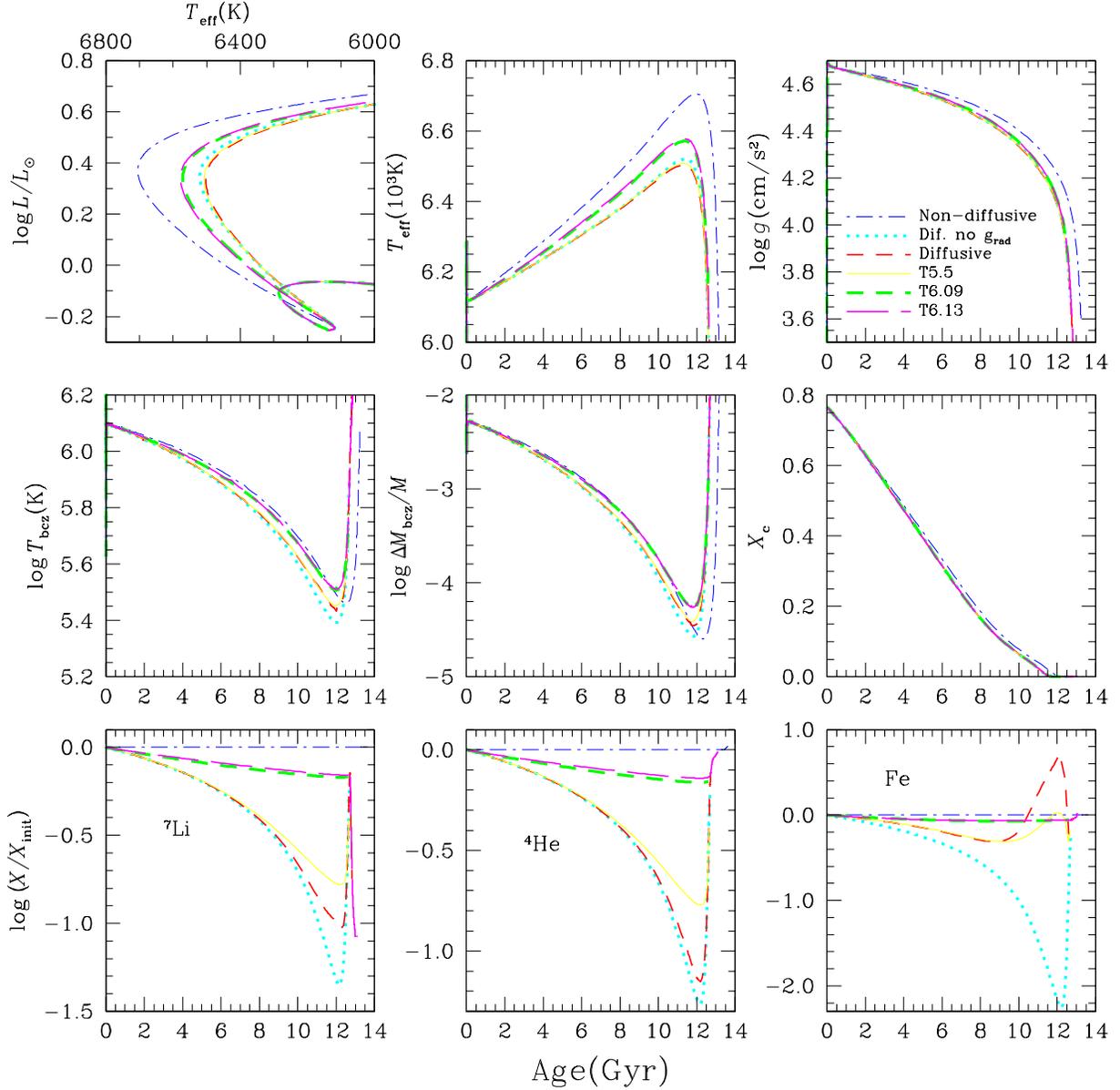}}
\caption{
Hertzsprung-Russell diagram, and time evolution of \teff{}, $\log g$, temperature at the base of the surface convection zone (\tbcz), 
mass at the base of the surface convection zone ({\ensuremath{M_{\scrbox{bcz}}}}) and mass fraction of hydrogen ($X_c$) at the  center 
and of the He, Fe and Li abundances in 0.8 \Msol{} models 
both with and without diffusion.  See Figure~\ref{fig:coefficients} and Table~\ref{tab:parameters} for a definition 
of the notation used for models with  turbulent transport. }

\label{fig:with_without}
\end{figure}

\begin{figure}
\centerline{
\includegraphics[width=\textwidth]{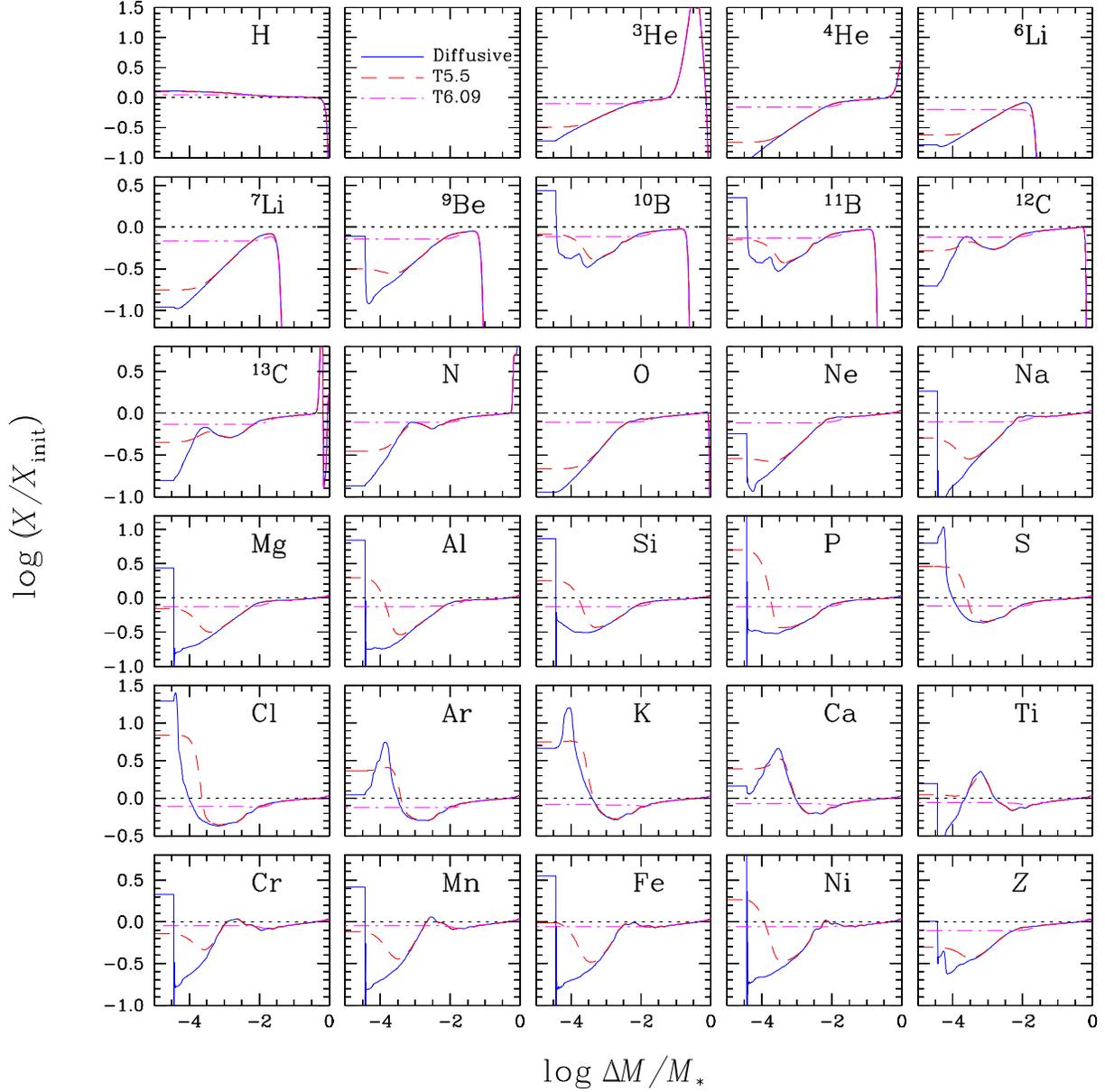}}
\caption{
Internal  abundance profiles  in a Pop II star of 0.8 \Msol{} with $[\Fe/\H]=-2.31$ and turbulence parametrized by T6.09D400-3 and  T5.5D400-3. The model with no turbulence is also shown for comparison purposes.
The profiles are shown at 11.7 Gyr, shortly before the star moves to the giant branch. The same scale is used as for Figure~\ref{fig:intern_abundances}.
  See Figure~\ref{fig:coefficients} and Table~\ref{tab:parameters} for a definition 
of the notation used for models with  turbulent transport. }
\label{fig:abtot_dif_6.09}
\end{figure}

\begin{figure}
\centerline{
\includegraphics[width=\textwidth]{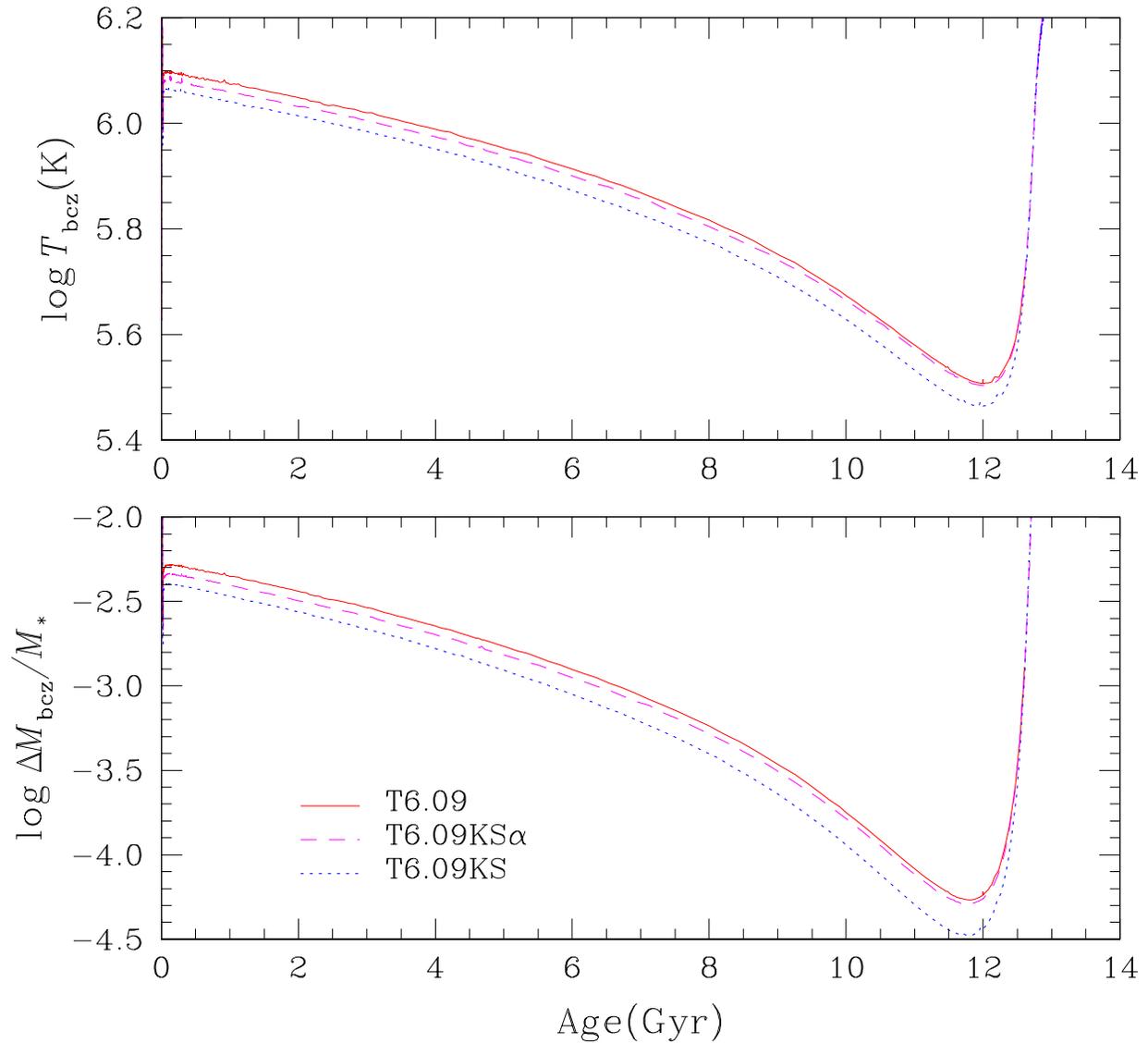}}
\caption{
Mass in the surface convection zone and temperature at its bottom as a function of  evolutionary time for a 0.8 \Msol{} model.
The models differ by the boundary conditions and by the value of $\alpha$ used in the calculations.
 See Figure~\ref{fig:coefficients} and Table~\ref{tab:parameters} for a definition 
of the notation used for models with  turbulent transport.
}
\label{fig:M_Tbzc}
\end{figure}

\begin{figure}
\centerline{
\includegraphics[width=0.45\textwidth]{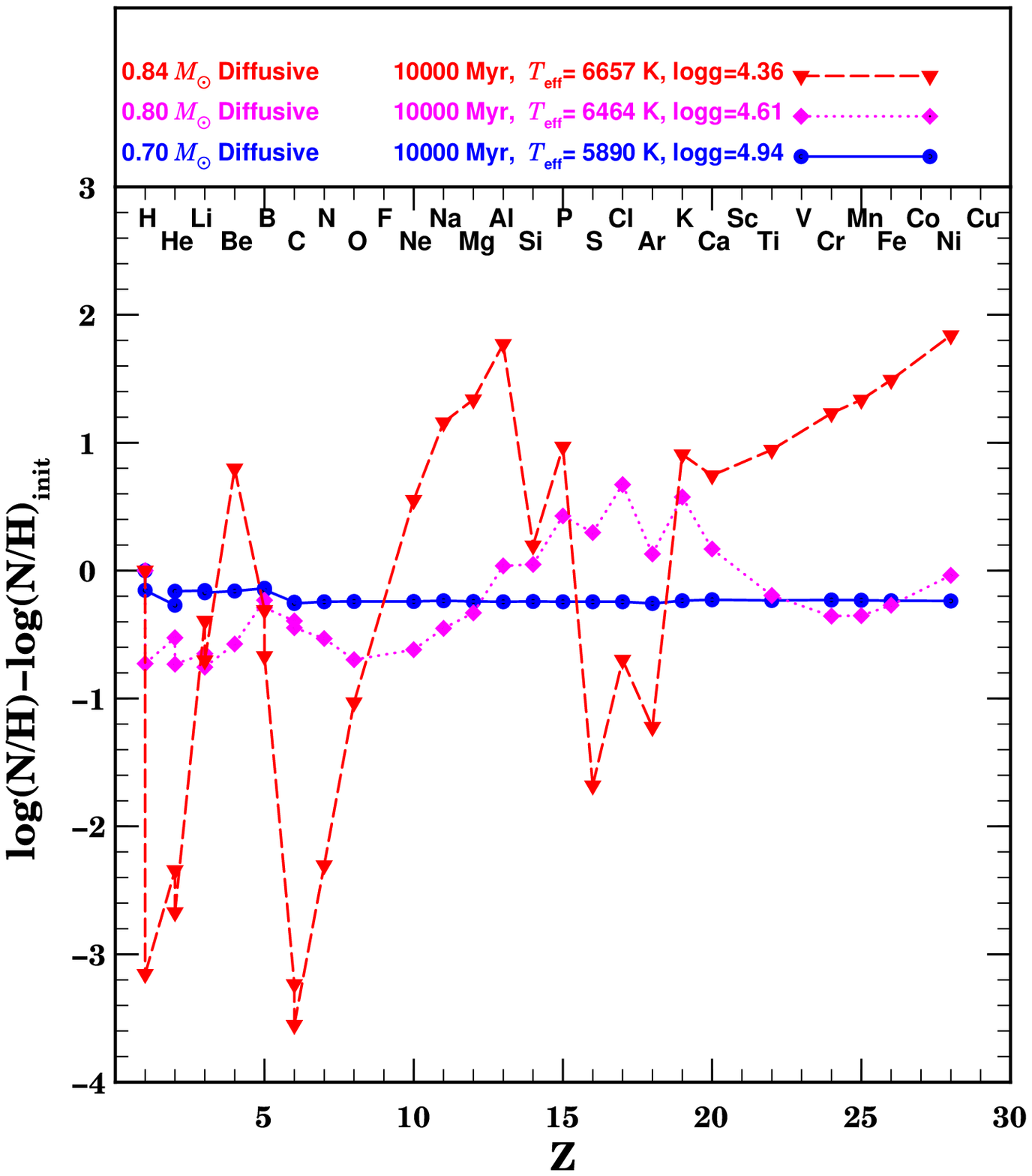}
\includegraphics[width=0.45\textwidth]{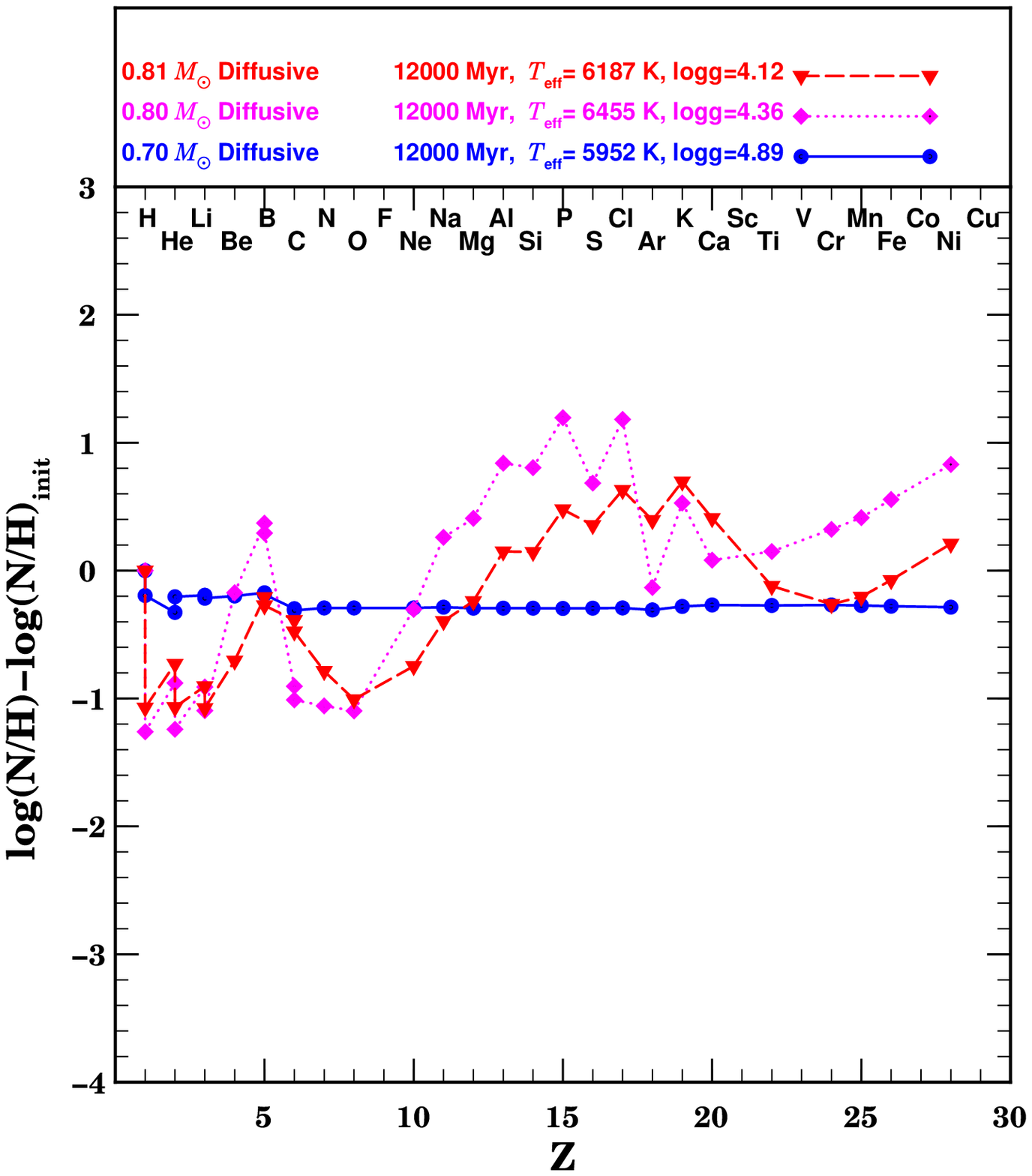}}
\caption{
In the left hand panel, ratio of surface abundances  to original abundances as a function of atomic mass 
in stars of 0.7, 0.8, and 0.84 \Msol{} after 10 Gyr of evolution.  In the right hand panel, the 0.84 \Msol{}
 model is replaced by the 0.81 \Msol{} one and it is after 12 Gyr of evolution.}
\label{fig:ab_Z10}
\end{figure}

The surface concentrations at 10 and 12 Gyr are shown as a function atomic number in Figure~\ref{fig:ab_Z10}
 for the evolutionary models with atomic diffusion and  no turbulence (see also 
subsection~\ref{sec:with diffusion}).  The 0.7 \Msol{} model is the one with the smallest mass shown.  In it,
 gravitational settling causes a general reduction of surface abundances by close to 0.2 dex.  As may be seen from 
Figure~\ref{fig:historic_noturb}, the convection zone is always at least 1\% of the  mass in a 0.7 \Msol{} star
 so that radiative 
accelerations play a modest  role (see Fig.~\ref{fig:grad}).  In the 0.6 and 0.5 \Msol{} models (not shown), the 
mass in the surface convection zone is always larger than about 20\,\% of the stellar mass so that gravitational 
settling is negligible even after 10 or 12 Gyr.  

In  stars of  0.75 \Msol{} or more, the surface convection zone occupies less than 0.1\,\% of the stellar mass
 for part of the evolution. The \gr{} then play the major role for that part of the evolution.  At 10 Gyr, the 
0.8 \Msol{} model is  significantly affected by \gr{} (see Fig.~\ref{fig:ab_Z10}). The atomic species most affected are 
those between Al and Ca.  They have overabundances by factors of up to 5.  The Fe peak elements are hardly affected.
In the 0.84 \Msol{} model, Ni is overabundant by a factor of 70, while Fe is overabundant 
by a factor of 30.  Large underabundances are expected for CNO and Li while Be and B are supported at least
 partly by their \gr.  Since the original metallic abundances are small, the \gr{} are little affected by line saturation 
and the abundance anomalies reflect essentially the atomic configuration of the dominant ionization state of each chemical species.  
For instance, Li is completely stripped of its electrons while C, N and O are mainly in  He-like configurations (see also \citealt{RicherMiRoetal98}).

At 12 Gyr, the 0.84 \Msol{} model has already evolved to the giant phase.  The most massive star 
around the turnoff has 0.81 \Msol{}.  It is already 300 K cooler than the turnoff so that its surface convection zone has
already started to expand and its abundance anomalies to decrease.  While its Fe peak shows approximately  the 
original abundances, 
the species between Al and Ca have  overabundances by up to a factor of 5.  Note again that Be and B 
are supported by their \gr{} while Li has sunk.  At that age, the 0.8 \Msol{} star is the star with the largest anomalies.
Ni and species between Na and Cl are about a factor of 15 overabundant.  He, Li and CNO are underabundant by a 
factor of about 10.   At 12 Gyr, the 0.80 \Msol{} is at the peak of its abundance anomalies: its
 surface convection zone is just about to start getting deeper (see Fig.~\ref{fig:historic_noturb}).   The 0.7 \Msol{} 
star also has larger anomalies than at 10 Gyr.  Its surface convection zone 
is still getting smaller (see Fig.~\ref{fig:historic_noturb}) while the 20\,\% longer life means more
time for gravitational settling.

\begin{figure}
\centerline{
\includegraphics[width=0.6\textwidth]{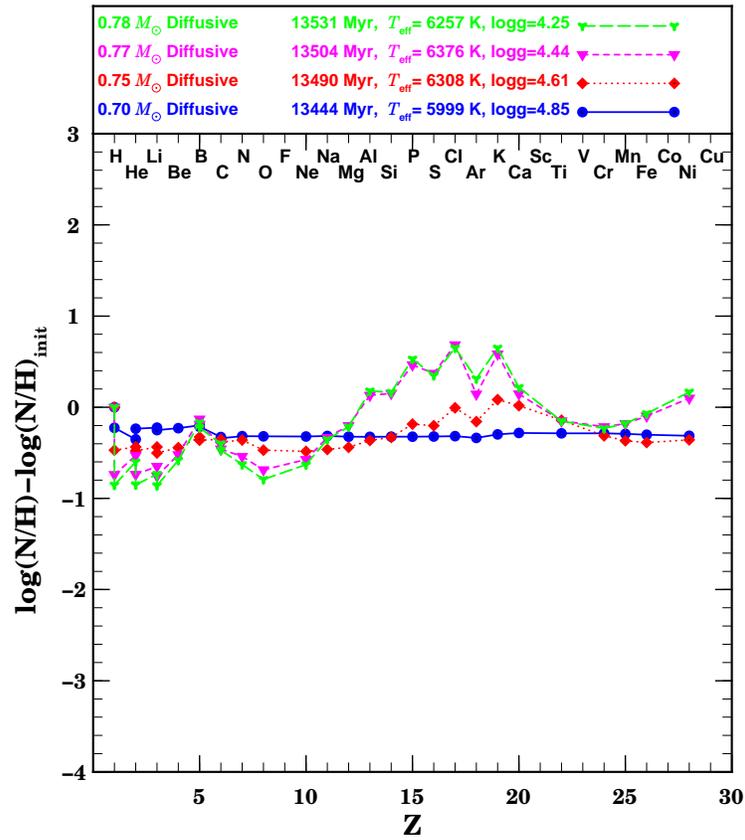}}
\caption{Ratio of surface abundances  to original abundances as a function of atomic mass,
 after 13.5 Gyr of evolution, in  0.7, 0.75, 0.77 and 0.78 \Msol{} stars.  
}
\label{fig:ab_Z13.5}
\end{figure}

At  13.5 Gyr (the age of the oldest clusters as determined in Paper II) 
the surface abundances are shown in Figure~\ref{fig:ab_Z13.5} for stars with no 
turbulence.  The lower mass star, 0.7 \Msol{}, is at $\teff = 6000$ K and has seen its 
original abundances reduced 
 by approximately 0.3 dex.  This applies, in particular, to Li. The slightly warmer (6300 K) 0.75 \Msol{} 
star has a larger Li underabundance ($-$0.5 dex), marginally larger Fe peak underabundances ($-$0.39 dex) but the 
Ca abundance goes back to its original value because it is supported by its \gr.  Chemical species between Al and
Ti are at least partially supported by their \gr.  The slightly more evolved 0.77 and 0.78 \Msol{} stars have nearly the 
same \teff{} but slightly lower gravities,  less massive surface convection zones (see Fig.~\ref{fig:historic_noturb}) and larger abundance anomalies than the 0.75 \Msol{} star.  The two have very similar 
abundance anomalies. Fe has about its original abundance while species between Al and Ca have larger surface abundances than their original ones.
Li is  underabundant by a factor of 7.3.  A comparison to  Figure~\ref{fig:ab_Z10} shows that, for instance, Ni and Al abundance anomalies are 
reduced by a factor of 10 as one goes from clusters of 10 Gyr to clusters of 12 Gyr and then by a factor of 5 as one goes from clusters of 12 Gyr to 
clusters of 13.5 Gyr.  As clusters age, the abundance anomalies caused by atomic diffusion in turnoff stars get progressively smaller.

\begin{figure}
\centerline{
\includegraphics[width=0.45\textwidth]{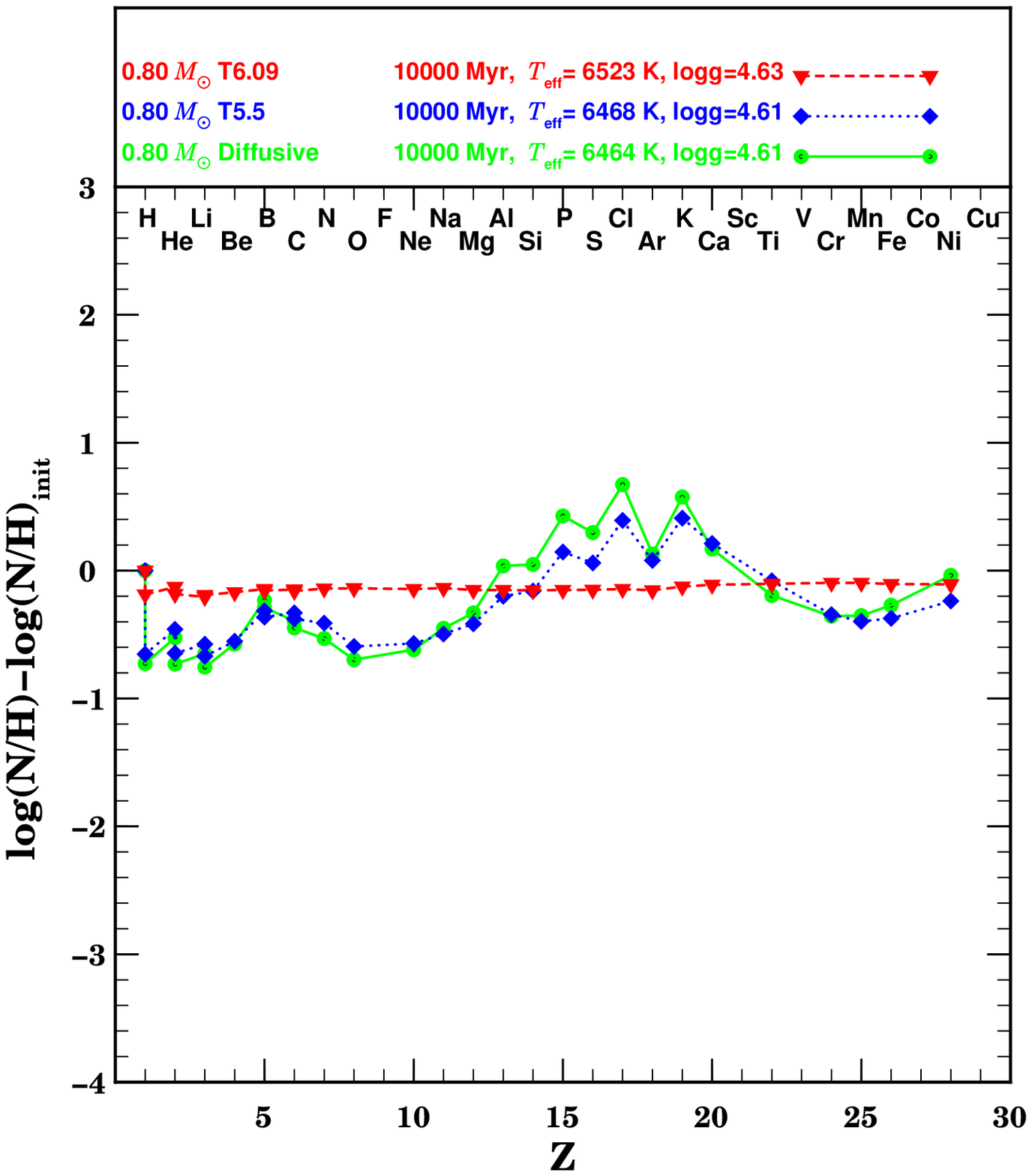}
\includegraphics[width=0.45\textwidth]{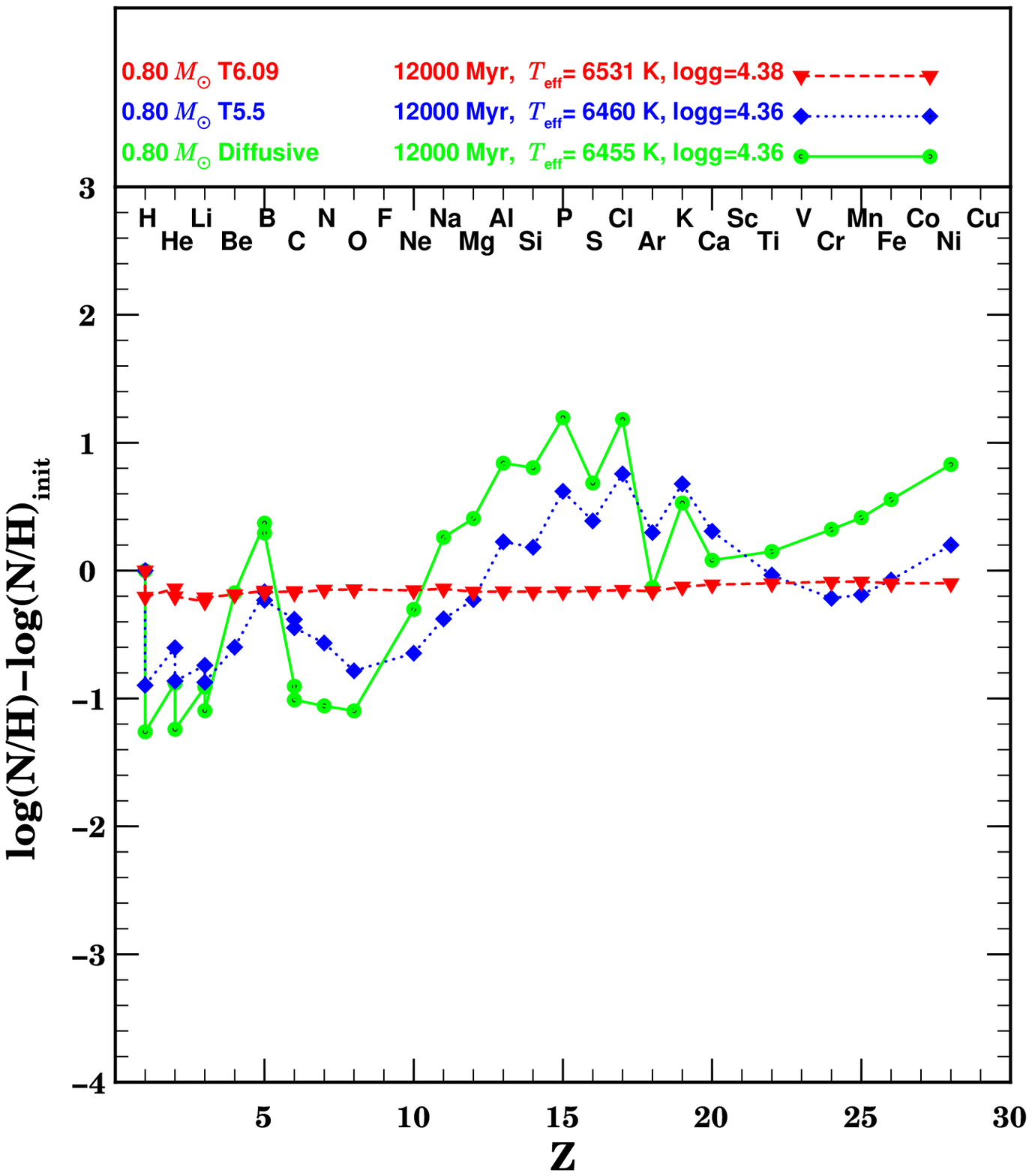}}
\caption{In the left/right hand panel, ratio of surface abundances  to original abundances as a function of atomic mass 
in a  0.8 \Msol{} star, after 10/12 Gyr of evolution for 3 turbulence strengths: atomic diffusion only, the T5.5 and 
T6.09 turbulence models.  
}
\label{fig:ab_Z12_08}
\end{figure}

The effect of turbulence strength on surface composition is shown for the 0.8 \Msol{} model in  
Figure~\ref{fig:ab_Z12_08} at 
10 and 12 Gyr.  As a general rule, turbulence reduces abundance anomalies.  However, in a 0.8 \Msol{} star, the 
effect of the T5.5D400-3 turbulence is more complex. 
Mg is mildly overabundant in the absence of 
turbulence but normal for the turbulence used, while Na goes from slightly over to slightly underabundant and the overabundance 
of Ca is increased by the increased turbulence.  The anomalies
between Al and Cl are more robust.  These differences in behavior may be related to the differences in the internal
distribution of the abundance variations caused  by diffusion for various species (see Fig.~\ref{fig:abtot_dif_6.09}).
At 11.7 Gyr, species from Na to Cl have their maximum concentrations in the surface convection zone in the absence of 
turbulence.  They are reduced by the extra mixing brought about by turbulence.  However, Ar, K, Ca and Ti have their
maximum concentrations below the convection zone so that some extra mixing increases their 
surface abundances.   As a comparison with Figure~\ref{fig:ab_Z10} shows, the effect of adding turbulence is 
 similar to that of considering a slightly more evolved, slightly more  massive model (compare the 0.8 \Msol{} star with T5.5  
of Fig.~\ref{fig:ab_Z12_08} with 
 the 0.81 \Msol{} of Fig.~\ref{fig:ab_Z10}),  
in that Fe peak anomalies are nearly eliminated in both while those 
between C and Ca are mildly reduced in both. The 0.78 \Msol{} no-turbulence model (not shown)  has very
 similar surface concentrations as both the 0.81 \Msol{} model without turbulence and the 0.8 \Msol{} star with T5.5 turbulence.
It differentiates itself  by slightly larger N, O and F underabundances.

However, the mixing 
that is strong enough to minimize Li abundance variations (the T6.09 model) goes deeper than the region 
where \gr{} play a substantial role. 
Turbulence then reduces the abundance anomalies of all
metals to 0.1$-$0.2 dex.  All metals are underabundant if mixing is that strong.

\subsection{As a function of $\teff$.}
\label{sec:ab_teff}
The 12 Gyr surface abundance isochrones are shown as a function of $\teff$ for a number of species 
($^6$Li, $^7$Li, Be, B, C, O, Na, 
Mg, Al, Si, S, Cl, Ca, Cr, Fe and Ni) in Figure~\ref{fig:ab_teff12}
 for the cases of  atomic diffusion with no turbulence  as well as with  two turbulence strengths (T5.5 and T6.09).
One first notes that, at the cool end, below 5000 K, there are only very small underabundances 
($\leq 0.1$ dex) for all species and all turbulence strengths, except
for Li which is destroyed by nuclear reactions and, to a lesser extent, Be.  As one considers hotter stars, 
all underabundance
factors increase until one reaches 6000 K.  There the \gr{} start to play a role for some species.  Furthermore,  
in the vicinity of the turnoff,  stars of similar \teff{} have very different evolution states.  In the absence of
turbulence, some turnoff stars can have overabundances of Fe by up to 0.7 dex. Around the turnoff
the Fe abundance can vary between  $-$0.3 dex underabundance and 0.7 dex overabundance.  
The presence of even a very small amount of turbulence  reduces this range by more than 0.7 dex 
(compare the continuous and dotted lines).   The  maximum abundance is
reduced to the original Fe abundance while the underabundance factor is only slightly reduced to 0.25 dex.  B, Mg, Cr, 
Mn and Ni are similarly affected by the same turbulence model though by slightly smaller factors.  However the overabundances between P (not shown) 
and Ca are more robust.  The range of Cl anomalies is merely reduced from a factor of 20 to a factor of about 7.  Similarly,
the large underabundances (He, Li, N and O; He and N are not shown) are not much reduced by the introduction of a small amount of turbulence.
In a given cluster, the range of chemical species calculated then offers the possibility of  testing turbulence models.

The surface abundance isochrones at the age determined for M92 in Paper II are shown in  Figure~\ref{fig:ab_teff13.5}.
  One first notes, by comparing to Figure~\ref{fig:ab_teff12}, the strong reduction in all  abundance variations
 around the turnoff but the general slight increase
in all underabundances in stars with $\teff \le 6000$ K.  The increased underabundances reflect the increased time available
 for gravitational settling and the slight reduction in the depth of the surface convection zone between 12 
and 13.5 Gyr (see Fig. \ref{fig:historic_noturb}).  

At 13.5 Gyr, one then expects, in the absence of turbulence, underabundances of 0.05 dex at $\teff{} = 5000$ K increasing to 0.1 dex around 5500 K.
Above that \teff{}, underabundances remain approximately constant for stars with the T6.09 turbulence model.  If there is no turbulence, 
underabundances increase to  0.3 dex at 6000 K and above that \teff{} both over and underabundances are present 
as \gr{} play a significant role.
For the Fe peak elements, abundances are between the original and 0.3 dex underabundance.  Only Ni could have an overabundance.  Between C 
and Na all species are underabundant by a factor of order 2 while species between Al and Ca are generally overabundant.  The presence of 
the T5.5  turbulence model mainly affects Al, Si, Fe and Ni overabundances which it reduces to or below the original abundance.  Its presence is
 sufficient to reduce the expected range  of underabundances of Fe to $-0.3 \ldots -0.1$ dex.

\section{Comparison to observations}
\label{sec:observations}
Since in this paper results are presented only for stars with $Z = 0.00017$, we will compare with 
the abundance determinations in M92 (see \citealt{BoesgaardDeStetal98} and \citealt{KingStBoetal98}).  In so far 
as one may assume  all cluster stars to share the same original composition, giant stars 
give a handle on the original turnoff star composition.  The observations of clusters with larger
$Z$ values will be discussed in a forthcoming paper where the evolutionary calculations for various
values of $Z$ will be presented.
We will use  Li abundance in field halo stars with $Z \leq 0.001$ to complement the M92 observations.
Since Li is believed to be of cosmological origin in these stars, one expects that the original abundance is the
 same for all. 

The $^7\Li$ observations of \citet{SpiteSp82} and  \citet{SpiteMaSp84}, who have established the presence of a Li 
abundance plateau, are compared in
 Figure~\ref{fig:Li_teff} to $^7\Li$ abundance isochrones at the age of M92 (13.5 Gyr) and at 12 Gyr. The observations
 of \citet{RyanNoBe99},  are also included since they further constrain the high $T$ part of the plateau.  Given these
results, it appears difficult to question the constraints imposed by the Li plateau, as done by \citet{SalarisWe2001}. 
 Results for 
3 series of models are shown: those with atomic diffusion without turbulence and those with the T5.5 and T6.09 
turbulent transport models.
The stars calculated with the T6.09 model clearly fit the $^7\Li$ observations much better.  Both at 12 and 13.5 Gyr, 
  there is a uniform  $^7\Li$ abundance extending from $\teff = 5500 $ to 6000 K, in agreement with observations. 
It corresponds to an underabundance by 0.17 dex from the original Li abundance.
Note also that $^6\Li$ is reduced by less than a factor of 2 in turnoff stars (see Fig.~\ref{fig:ab_teff12} and 
Fig.~\ref{fig:ab_teff13.5}), which is compatible with the observations 
of the presence of  $^6\Li$ in those Pop II objects  (see \citealt{CayrelSpSpetal99}).

The series of models calculated without turbulence and that calculated with the T5.5 model do not reproduce the 
observations for $\teff \geq 6100 $ K.  Given the observational error bars, the T6.09 model leads to a \teff{} dependence 
of $X$(Li) which is perhaps more constant than required.  However, at 12 Gyr,  calculations  with the 
T6.0 model (not shown) give a Li abundance 0.15 dex smaller at 6500 K than at 5600 K.  
While perhaps compatible with the error bars, this would lead to a significantly poorer fit and may be 
considered the lower limit to the turbulent transport required by observations.

The lithium abundances observed in the M92 turnoff  stars by \citet{BoesgaardDeStetal98} are also 
shown in the 13.5 Gyr  part of the figure.  The error bars are much larger than for field stars mainly 
because of the much smaller signal to noise ratio.    The determined Li abundances 
 suggest a slightly larger original value for the Li abundance in M92 than in halo stars but, given the size of 
the error bars, this could be considered uncertain.   One star has a Li abundance higher than the others by a factor larger
than the error bars.  Other stars at the same \teff{} have a clearly smaller equivalent width of the Li doublet.  
The authors used this to argue for Li destruction by a large factor in Pop II 
stars.  This appears premature to us.

  \citet{KingStBoetal98} have  observed 7 metals  mainly in three subgiant stars of M92 with $\teff \simeq 6020$ K  (see their
discussion of \teff{} in their section 4.3), 
  which is slightly cooler than our 6300 K turnoff \teff{} at 13.5 Gyr.  Their 
determination of the Fe abundance is based on more lines and appears more  reliable than that of other metals.
It is the only metal we discuss.  Their main conclusion is that the subgiants of M92 have a 0.26 dex lower Fe abundance
than the giants of M92 as determined by   \citet{SnedenKrPretal91}.  They carefully discuss all sources of error 
and cannot completely exclude that this 0.26 dex difference be reduced even possibly to zero.  If the T6.09 turbulent
 transport model applies, all 3 stars should have the same Fe 
abundance and it should be approximately 0.1 dex smaller than the original Fe abundance  according to
 our  Figure~\ref{fig:ab_teff13.5}.  The giants should have the original Fe abundance because of their large 
surface convection zones. This seems quite compatible 
with the observations of Fe given the uncertainties.

 \citet{KingStBoetal98} also find (their 
section 4.2) that star 21 has a 0.15 dex larger Fe abundance than the other two (stars 18 and 46).
 Furthermore  \citet{BoesgaardDeStetal98}
observed that the $^7$Li abundance is a factor of 3 smaller in star 21 than in star 18.  This is to be compared to our results
in  Figure~\ref{fig:ab_teff13.5}.  At the turnoff, the Li and Fe abundances are sensitive functions of both the exact evolutionary  state and turbulence.
At 13.5 Gyr, in 0.78 \Msol{} models without turbulence, the Fe abundance is $-$0.05 dex from normal.  Introducing the T5.5 turbulence 
model reduces it by 0.25 dex while the T6.09 model brings it back to $-$0.1 dex from normal.
As may be seen in Figure \ref{fig:LiFe_teff}, the Fe abundance as a function of \teff{} has a complex loop structure
 around the turnoff in both  0.8 and 0.78 \Msol{}
stars.  Small changes in evolutionary state and/or turbulence can change the Fe abundance by 0.2$-$0.3 dex and the $^7$Li abundance by 
0.5$-$0.6 dex.  At the turnoff, the different Li and Fe abundances 
would then be explained by a different turbulence between the stars, resolving  
the difficulty described in Section 4.2 of  \citet{KingStBoetal98}.   They would become the Pop II analogue of AmFm stars.  
However all three of those stars appear to
be past the turnoff.   As Figure \ref{fig:LiFe_teff} shows, the \gr{} have significant effects on surface abundances 
in stars of mass 0.77 \Msol{} and more but, for Fe, these effects are limited to $\teff \geq 6100 $ K. 
As seen in Figure \ref{fig:ab_Z_teff}, the abundance variations at the turnoff 
are not expected in subgiants when they reach 6020 K. At this temperature, reducing turbulence leads to more gravitational settling for both 
Li and Fe which cannot explain the differences between stars 18 and 21.
 While turnoff stars of a cluster could have  abundance 
variations as observed for stars 18, 21 and 46, these are not expected in stars with exactly the parameters determined
for them---and this is a difficulty even if  there appears to be considerable uncertainty in the 
\teff{} of stars in this cluster (compare the \teff{} of \citealt{Carney83} to those of  \citealt{King93} and the discussion 
in Section 4.3 of  \citealt{KingStBoetal98}).  Note that in the comparison to our evolutionary models, it is the absolute \teff{}
scale that matters and not just the \teff{} with respect to the turnoff \teff. If anomalies are confirmed,
 agreement with our model would require the higher \teff{} scale for M92.

\section{Conclusion}
\label{sec:conlusion}

It has been clearly shown that, contrary to the belief expressed in many  papers and recently by
 \citet{ChaboyerFeNeetal2001}, atomic diffusion does not necessarily lead to underabundances 
of metals in Pop II stars.  Differential radiative accelerations lead to  overabundances of Fe and some other
 chemical elements in some turnoff stars.

\begin{figure}
\centerline{
\includegraphics[width=\textwidth]{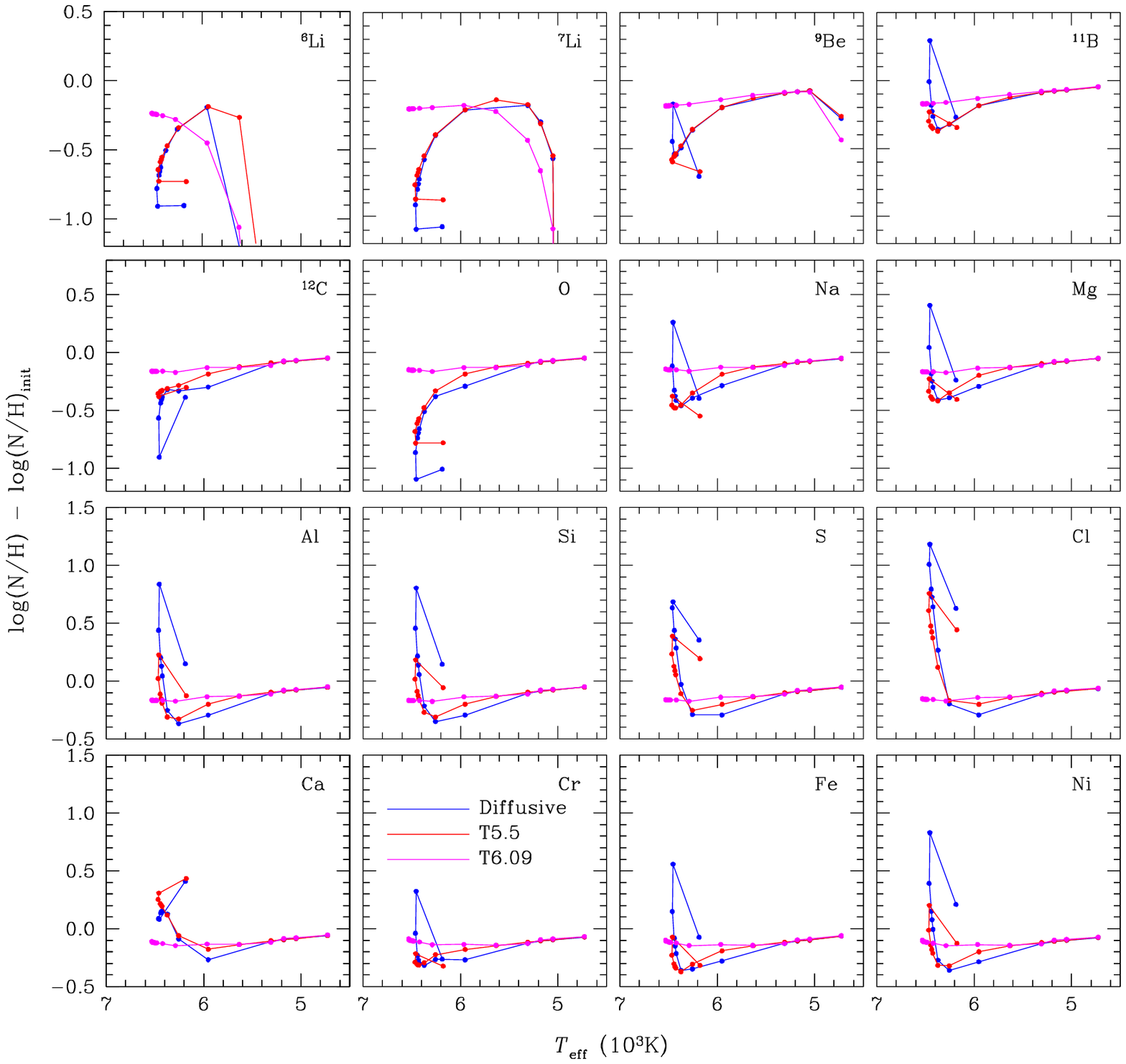}}
\caption{
Ratio of surface to original abundances  after 12 Gyr of evolution as a function of \teff. The continuous line
segments link models calculated with atomic diffusion.  The dashed and dotted line
segments link models calculated with atomic diffusion and, respectively, the T6.09 and T5.5 turbulence models.  
}
\label{fig:ab_teff12}
\end{figure}

\begin{figure}
\centerline{
\includegraphics[width=\textwidth]{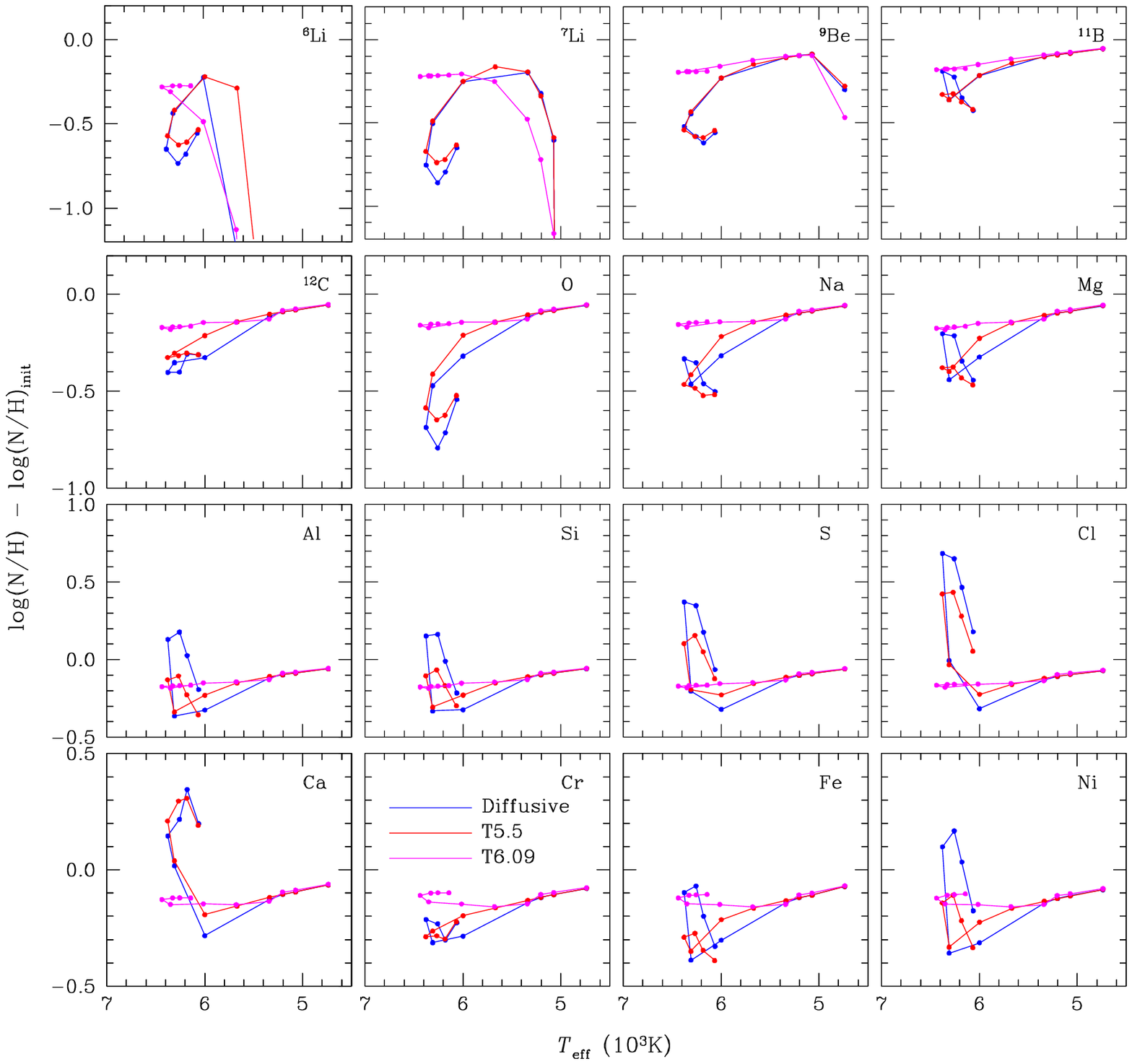}}
\caption{
Ratio of surface to original abundances  after 13.5 Gyr of evolution as a function of \teff. The continuous line
segments link models calculated with atomic diffusion.  The dashed and dotted line
segments link models calculated with atomic diffusion and, respectively, the T6.09 and T5.5 turbulence models.  }
\label{fig:ab_teff13.5}
\end{figure}

\begin{figure}
\centerline{
\includegraphics[width=\textwidth]{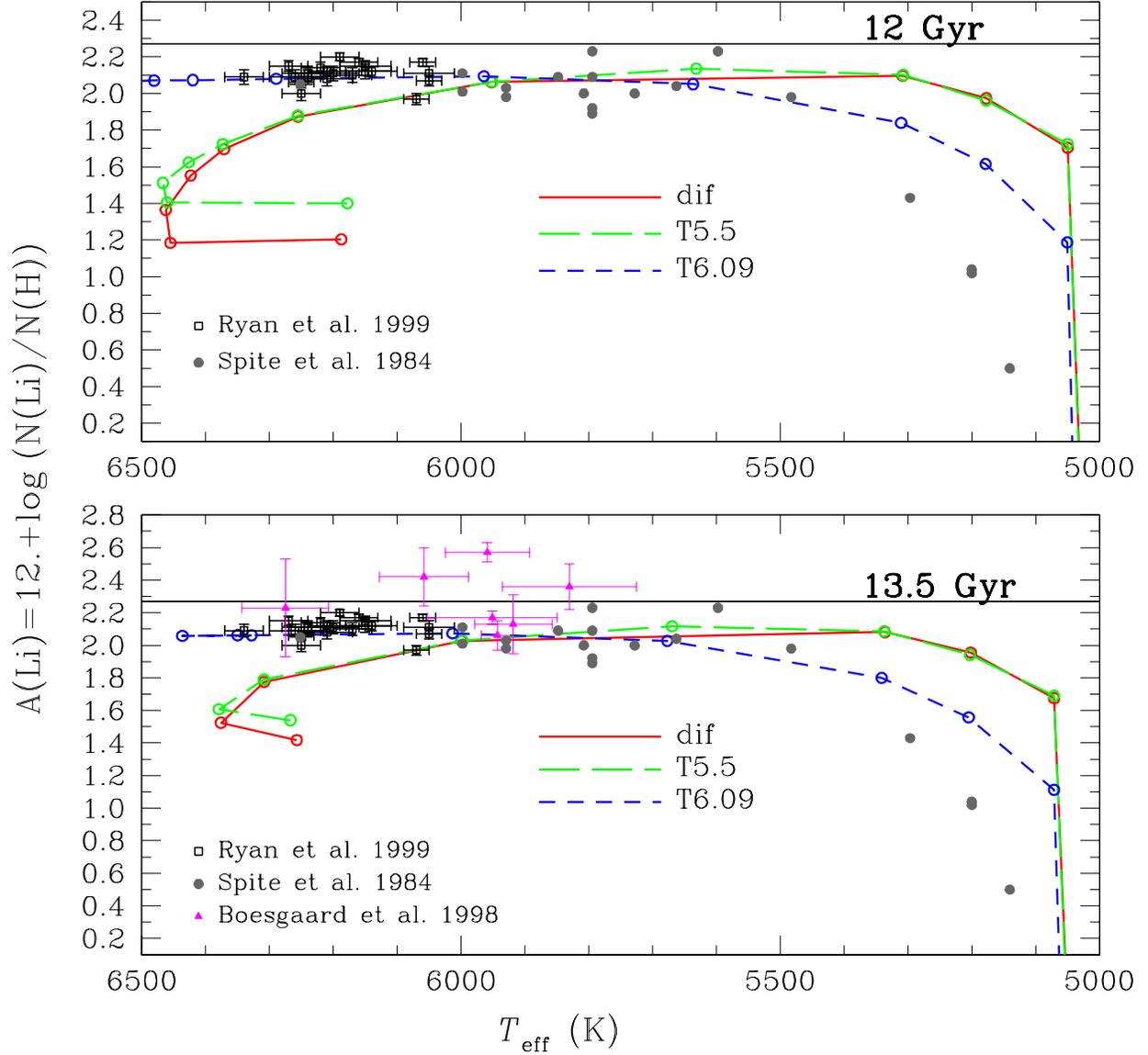}}
\caption{
Lithium surface abundances  after 12 and 13.5 Gyr of evolution for 3 turbulence strengths: no turbulence,
 similar to AmFm stars (T5.5) and minimizing Li surface abundance variations (T6.09).  Segmented 
straight lines link calculated values.  In comparing results
 for different turbulence models for $\teff \leq$ 6000 K, one should
remember that, at a given mass and age, a star with the T6.09 turbulence model has a higher \teff{} 
(by some 100 \Kelvin{}) than one
with atomic diffusion only (see Fig. \ref{fig:with_without}).
References for the observed Li abundances are identified on the figure.  }
\label{fig:Li_teff}
\end{figure}

\begin{figure}
\centerline{    
\includegraphics[width=\textwidth]{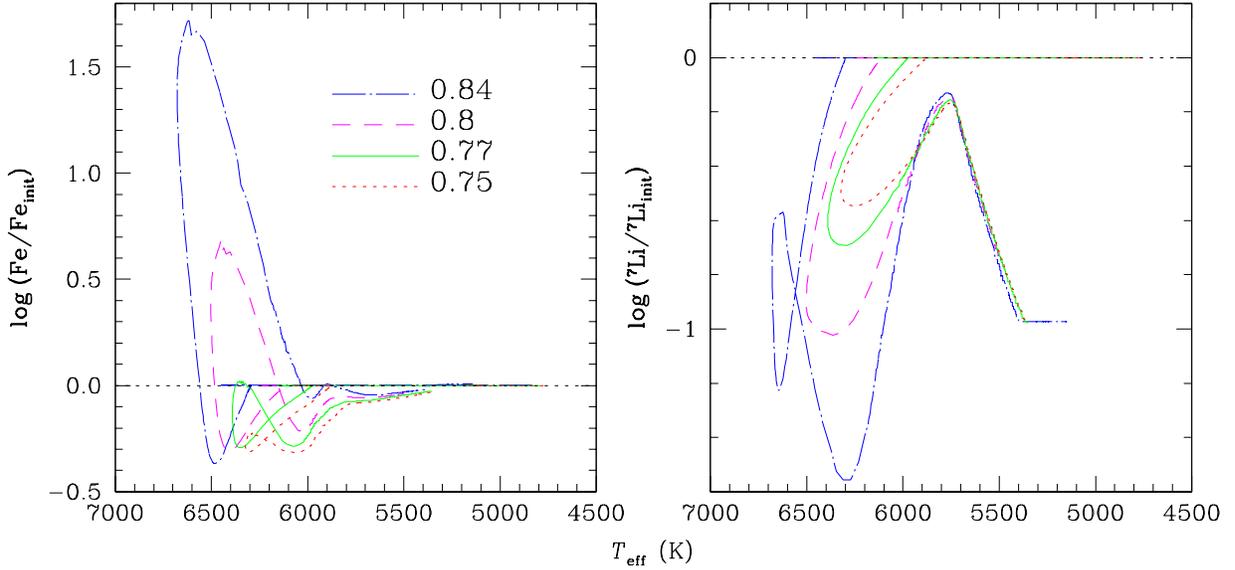}}
\caption{
Lithium and Fe surface abundances as a function of \teff{} throughout the main sequence and subgiant
evolution of 0.84, 0.8, 0.77 and 0.75 \Msol{} stars when atomic diffusion is included in the calculations.  
Turbulence is assumed negligible.  For the
0.84 \Msol{} star the main sequence starts at $T=6300 $K where  $\Fe =\Fe_0$.  As evolution proceeds, the
Fe abundance decreases until an underabundance of $-$0.4 dex is reached.  At that point, the surface convection
zone has retracted sufficiently for \gr{(\Fe)} to be greater than gravity and the Fe abundance increases up
to a 1.7 dex overabundance. This occurs at the end of the main sequence evolution (around the turnoff) at $\teff = 6600$K.
  As evolution proceeds on the subgiant branch the surface convection zone gets deeper and the Fe abundance 
decreases to 0.1 dex below the original abundance.  The Fe abundance goes back to its 
original value when a  \teff{} of 5300$-$5400 K is reached.  Similar 
evolution occurs for the other stars except that it starts at different \teff{} for each. The Li abundance in the 
0.84 \Msol{} star has a similar evolution.  The
Li abundance decreases until an underabundance of $-$1.2 dex is reached.  At that point, the surface convection
zone has retracted sufficiently for \gr{(\Li)} to be greater than gravity.  However \gr(\Li) is larger than gravity 
over a smaller interval than \gr{(\Fe)}.
  It leads to an increase of the Li abundance at $\teff = 6700 $K but to no overabundance. 
As the subgiant evolution starts,
 Li gets below its main sequence abundance at $\teff = 6500 $K, creating a loop in the Li abundance curve.  
The Li abundance never
 reaches its original abundance since, before this occurs, the surface convection zone approaches  the region of Li burning
and the Li abundance decreases rapidly below 5800 K  .}
\label{fig:LiFe_teff}
\end{figure}

\begin{figure}
\centerline{
\includegraphics[width=0.6\textwidth]{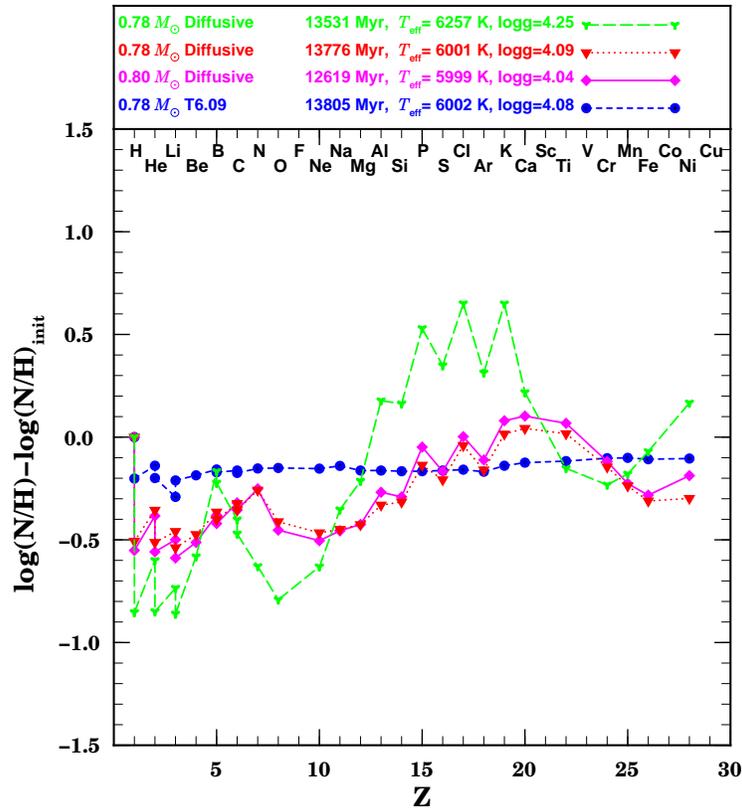}}
\caption{
Surface abundances  at $\teff = 6000 $K  for  a 0.8 and a 0.78 \Msol{} star with atomic diffusion only.  Their ages
bracket the age of M92, 13.5 Gyr.  Both stars have very closely the same composition showing that surface
abundances are not a sensitive function of mass at $\teff = 6000 $K, when the star is on the subgiant branch.
  A 0.78 \Msol{} star with the T6.09
turbulence model is also shown at the same \teff{}.  Finally the 0.78 \Msol{} star with no turbulence is shown at 13.5 Gyr
 when it is a turnoff star ($\teff = 6250 $K). }
\label{fig:ab_Z_teff}
\end{figure}

Consider the evolution of Pop II stars with no tubulence.  As one may see by considering the solid curves 
in Figures~\ref{fig:ab_teff12} and \ref{fig:ab_teff13.5}, generalized underabundances
by 0.1 dex are expected in the \teff{} interval from 4600 to 5500 \Kelvin{}.  Between 5500 and 6000 \Kelvin{}, the underabundances are still 
generalized and increase to 0.3 dex for some species such as \Ox{}.  In 12 Gyr turnoff stars however ($\teff{} \geq 6000$ \Kelvin{}), overabundances by a 
factor of up to 10 are possible (e.g. Al and Ni).  All calculated species heavier than Na may have overabundances. 
 At a given \teff{}, variations are expected from star to star.  
At 13.5 Gyr, similar but smaller anomalies are expected.  The overabundances are sensitive to any left over turbulence below the convection zone.

In this paper, the evolution of stars with $Z = 0.00017$ has been described both with and without turbulence.  A 0.1 dex underabundance of metals in
turnoff stars as compared to giants has been shown to be the smallest anomaly to be expected (section \ref{sec:ab_teff}).
   Star-- to-- star variations were seen to be
possible around the turnoff, if turbulence is small enough.
 Observations (see section \ref{sec:observations}) 
suggest the presence of abundance variations similar to those expected.  The comparison to observations 
is, however, sensitive to the \teff{} scale.   As \citet{KingStBoetal98} concluded at the end of their section 4.2,
 higher quality data is
 probably required to establish the reality of Fe abundance 
variations within M92. The accurate determination of the abundance of more species is also needed.  This may well have 
implications not only for intrinsic abundance variations in clusters but for  
internal stellar structure.

The effect of varying $Z$ on the evolution of Pop II stars will be investigated in a forthcoming paper, 
(Paper III), where comparisons
 to higher $Z$ clusters will be undertaken. Increasing $Z$ in Pop II stars will be shown to reduce considerably the expected abundance
 anomalies.  Note also that Paper II shows that the present models for $[\Fe/\H]_0 = -1.31$ accurately
reproduce the CMD locations of local
Population II subdwarfs having Hipparcos parallaxes and metallicities within
+/$-$ 0.2 dex of $[\Fe/\H] = -1.3$. 

In a number of
clusters with higher $Z$ than M92, observations suggest that only small variations, if any,  are present in  turnoff stars 
(see for instance  \citealt{RamirezCoBuetal2001};
 \citealt{GrattonBoBretal2001} and \citealt{TheveninChdeetal2001}).  Furthermore in field halo stars, the 
Li abundance puts strict constraints on any chemical separation.  In the companion paper (Paper II) we 
therefore took the cautious approach to use mainly evolutionary models that minimize the effect of atomic diffusion.

It has been shown that the use, in complete stellar evolution models, of a relatively simple parametrization of 
turbulent transport leads to Li surface abundances
compatible with  the Li plateau observed in field halo
stars (with a 0.17  dex reduction from the original Li abundance) and small variations in the surface 
abundances of metals (a 0.1 dex reduction of the metal abundance in turnoff stars as compared to that in giants 
in clusters with $Z = 0.00017$).
  At the same time, the gravitational 
settling of He leads to a reduction in the age of
 globular clusters by some 10\,\% (see Paper II).
However simple the parametrization of turbulent transport, it is not understood from first principles.  
The high level of constancy
of Li abundance as a function of \teff{} requires that  turbulence mixes to very nearly the
same $T$ throughout the star evolution and in stars covering  a mass interval of approximately 0.6 to 0.8 \Msol{}.  
As already noted by 
\citet{MichaudFoBe84}, this is not expected in standard stellar models.  No convincing hydrodynamic model has been 
proposed that explains this property.  Pop II stars appear to tell us that this is the case, however.  

Mass loss is another physical process that could compete with atomic diffusion and maintain a constant value 
of Li as a function of \teff{} \citep{VauclairCh95}.   Whether, in the absence 
of turbulent transport,  it could be made consistent with
the observations of metals in globular cluster turnoff stars is a question that requires further calculations. 
These may lead to observational tests of the relative importance of mass loss and turbulence
in these objects.  The number of chemical species that are now included in these calculations and that can be
observed makes such tests possible.

\acknowledgments
We thank an anonymous referee for constructive comments that led to significant improvements of this paper. 
This research was partially supported at the University of Victoria and the Universit\'e de Montr\'eal
by
NSERC. We thank the R\'eseau Qu\'eb\'ecois de Calcul de Haute
Performance (RQCHP)
for providing us with the computational resources required for this
work.  S.T. extends his warm thanks to the Service d'Astrophysique at CEA-Saclay
 for an enjoyable and productive stay during which part of this research
 was performed.
 This work was performed in part under the auspices of the U.S.
 Department of Energy, National Nuclear Security Administration, by the
 University of California, Lawrence Livermore National Laboratory under
 contract No.W-7405-Eng-48.

\newpage

\clearpage

\begin{deluxetable}{lccc}
\tabletypesize{\scriptsize}
\tablewidth{.45\textwidth}
\tablecaption{Initial chemical composition}
\tablecolumns{4}
\tablehead{\colhead{Element} & {\tiny Note} & \multicolumn{2}{c}{Mass fraction} \\
\cline{3-4} & & [Fe/H]=-2.31 & [Fe/H]=-1.31 }
\startdata
H  \dotfill & & 7.646$\times 10^{-1}$ & 7.613$\times 10^{-1}$ \\
$^4$He \dotfill & \tablenotemark{a} & 2.352$\times 10^{-1}$ & 2.370$\times 10^{-1}$ \\
$^{12}$C  \dotfill & \tablenotemark{b} & 1.727$\times 10^{-5}$ & 1.727$\times 10^{-4}$ \\
N  \dotfill & & 5.294$\times 10^{-6}$ & 5.294$\times 10^{-5}$ \\
O  \dotfill & \tablenotemark{c} & 9.612$\times 10^{-5}$ & 9.612$\times 10^{-4}$ \\
Ne \dotfill & \tablenotemark{c} & 1.966$\times 10^{-5}$ & 1.966$\times 10^{-4}$ \\
Na \dotfill & \tablenotemark{c} & 3.986$\times 10^{-7}$ & 3.986$\times 10^{-6}$ \\
Mg \dotfill & \tablenotemark{c} & 7.484$\times 10^{-6}$ & 7.484$\times 10^{-5}$ \\
Al \dotfill & \tablenotemark{d} & 1.627$\times 10^{-7}$ & 1.627$\times 10^{-6}$ \\
Si \dotfill & \tablenotemark{c} & 8.072$\times 10^{-6}$ & 8.072$\times 10^{-5}$ \\
P  \dotfill & \tablenotemark{c}  & 6.976$\times 10^{-8}$ & 6.976$\times 10^{-7}$ \\
S  \dotfill & \tablenotemark{c}  & 4.215$\times 10^{-6}$ & 4.215$\times 10^{-5}$ \\
Cl \dotfill & \tablenotemark{c} & 8.969$\times 10^{-8}$ & 8.969$\times 10^{-7}$ \\
Ar \dotfill & \tablenotemark{c} & 1.076$\times 10^{-6}$ & 1.076$\times 10^{-5}$ \\
K  \dotfill & \tablenotemark{c}  & 1.998$\times 10^{-8}$ & 1.998$\times 10^{-7}$ \\
Ca \dotfill & \tablenotemark{c} & 7.474$\times 10^{-7}$ & 7.474$\times 10^{-6}$ \\
Ti \dotfill & \tablenotemark{c} & 3.986$\times 10^{-8}$ & 3.986$\times 10^{-7}$ \\
Cr \dotfill & & 9.989$\times 10^{-8}$ & 9.989$\times 10^{-7}$ \\
Mn \dotfill & \tablenotemark{e} & 3.890$\times 10^{-8}$ & 3.890$\times 10^{-7}$ \\
Fe \dotfill & & 7.172$\times 10^{-6}$ & 7.172$\times 10^{-5}$ \\
Ni \dotfill & & 4.445$\times 10^{-7}$ & 4.445$\times 10^{-6}$ \\
\cline{1-4}
Z  \dotfill & & 1.675$\times 10^{-4}$ & 1.675$\times 10^{-3}$
\enddata
\tablenotetext{a}{\;$^3\He=5.000 \times 10^{-5}$}
\tablenotetext{b}{\;$^{13}$C is 1\% of $^{12}$C}
\tablenotetext{c}{\;$[{\rm X}/\Fe]=+0.3$ (see \citealp{VandenBergSwRoetal2000})}
\tablenotetext{d}{\;$[{\rm X}/\Fe]=-0.3$ (see \citealp{VandenBergSwRoetal2000})}
\tablenotetext{e}{\;$[{\rm X}/\Fe]=-0.15$ (see \citealp{VandenBergSwRoetal2000})}
\label{tab:Xinit}
\end{deluxetable}

\clearpage

\begin{deluxetable}{lccccc}
\tablecaption{Parameters used for the different series of models identified in the first column }
\tablecolumns{6}
\tablehead{ \colhead{Series name} & \colhead{Boundary condition} &
\colhead{$\alpha{}$} & \colhead{[Fe/H]} & \colhead{Diffusion and} &
\colhead{Turbulence\tablenotemark{a}} \\
 & & & & radiative accelerations & }
\startdata
Non-diffusive  & Eddington & 1.687 & $-2.31$ & no  & no \\
Diffusive   & Eddington & 1.687 & $-2.31$ & yes & no \\
T5.5   & Eddington & 1.687 & $-2.31$ & yes & T5.5D400-3 \\
T6.0   & Eddington & 1.687 & $-2.31$ & yes & T6.0D400-3 \\
T6.09 & Eddington & 1.687 & $-2.31$ & yes & T6.09D400-3 \\
T6.13  & Eddington & 1.687 & $-2.31$ & yes & T6.13D400-3 \\
T6.09KS & Krishna-Swamy & 1.869 & $-2.31$ & yes & T6.09D400-3 \\
T6.09KS$\alpha$  & Krishna-Swamy & 2.017 & $-2.31$ & yes & T6.09D400-3
\\
Non-diffusive$-1.31$  & Eddington & 1.687 & $-1.31$ & no  & no \\
Diffusive$-1.31$    & Eddington & 1.687 & $-1.31$ & yes & no \\
T6.0$-1.31$   & Eddington & 1.687 & $-1.31$ & yes & T6.00D400-3 \\
T6.09$-1.31$  & Eddington & 1.687 & $-1.31$ & yes & T6.09D400-3 \\
\enddata
\tablenotetext{a}{\;see section~\ref{subsec:model}}
\label{tab:parameters}
\end{deluxetable}

\end{document}